\documentstyle[12pt]{article}
\newcommand\be{\begin{equation}}
\newcommand\ee{\end{equation}}
\newcommand\bea{\begin{eqnarray}}
\newcommand\eea{\end{eqnarray}}

\newcommand{\fatalpha}{{\bf \alpha \kern -0.44em \alpha}}
\newcommand{\fatsigma}{{\bf \sigma \kern -0.54em \sigma}}
\newcommand{\tpchi}{{\bf \chi \kern -0.35em \chi}}
\newcommand{\llambda}{{\bf \lambda \kern -0.45em \lambda}}

\bibliography{plain}
\bibliography{plain}
\title{\bf Investigating graph isomorphism in cospectral graphs via multiparticle quantum walk in fermionic basis and entanglement entropy }\vspace{20mm}
\author{ M. A. Jafarizadeh$^{a}$
 \thanks{E-mail:jafarizadeh@tabrizu.ac.ir},
 F. Eghbalifam$^{a}$
 \thanks{E-mail:F.Egbali@tabrizu.ac.ir},
 S. Nami$^{a}$
 \thanks{E-mail:S.Nami@tabrizu.ac.ir}
\\ $^a${\small Department of Theoretical Physics and Astrophysics,
University of Tabriz, Tabriz 51664, Iran.}} \pagebreak

\vspace{20mm}
\begin{document}
\maketitle \vspace{15mm}

\begin{abstract}
We investigate the graph isomorphism (GI) in some cospectral networks. Two graph are isomorphic when they are related to each
other by a relabeling of the graph vertices. We want to investigate the GI in two scalable (n + 2)-regular graphs $G_4(n; n + 2)$ and $G_5(n; n + 2)$, analytically by using the multiparticle quantum walk. These two graphs are a pair of non-isomorphic connected cospectral regular graphs for any positive integer $n$. In order to investigation GI in these two graphs, we rewrite the adjacency matrices of graphs in the antisymmetric fermionic basis and show that they are different for these pairs of graphs. So the multiparticle quantum walk is able to distinguish pairs of non-isomorph graphs. Also we construct two new graphs $T_4(n;n+2)$ and $T_5(n;n+2)$ and repeat the same process of $G_4$ and $G_5$ to study the GI problem by using multiparticle quantum walk. Then we study GI by using the entanglement entropy. To this aim, we calculate entanglement entropy between two parts of network. In our model the nodes are considered as identical quantum harmonic oscillators. The entanglement entropy between two special parts of $G_4(n; n + 2)$ and $G_5(n; n + 2)$ are calculated analytically. It is shown that the entanglement entropy can distinguish pairs of non-isomorphic cospectral graphs too.
\end{abstract}

\newpage
\section{Introduction}
One of the important problems about networks is the graph isomorphism (GI)
problem [1]. Two graphs are isomorphic, if one can be transformed into
the other by a relabeling of vertices (i.e. two graphs with the
same number of vertices and edges are nonisomorph, if they can not
be transformed into each other by relabeling of vertices). Many
graph pairs may be distinguished by a classical algorithm which
runs in a time polynomial in the number of vertices of the graphs,
but there exist pairs which are computationally difficult to
distinguish. Currently, the best general classical algorithm has a
run time $O(c^{\sqrt{N}\log N})$, where $c$ is a constant and $N$ is the
number of vertices in the two graphs.
Typical instances of graph isomorphism (GI) can be solved in polynomial time because two randomly chosen graphs with identical numbers of vertices and edges typically have different degree and eigenvalue distributions. Moreover, GI can be solved efficiently
for restricted classes of graphs, such as trees[2], planar graphs[3], graphs with bounded
degree[4], bounded eigenvalue multiplicity[5], and bounded average genus[6].
Researchers have also recently attacked GI using various
methods inspired by physical systems. Rudolph mapped the
GI problem onto a system of hard-core atoms [7].
Gudkov and Nussinov proposed a
physically motivated classical algorithm to distinguish nonisomorphic graphs [8].

Some researchers used quantum random walks (QRW) to test the capability
of quantum walks to distinguish nonisomorphic graphs.
Shiau et al. proved that the simplest
classical algorithm fails to distinguish some pairs of nonisomorphic graphs and also proved that continuous-time
one-particle QRWs cannot distinguish some non-isomorphic
graphs [9]. Douglas and Wang modified a single-particle
QRW by adding phase inhomogeneities, altering the evolution
as the particle walked through the graph [10].
Emms et al. used discrete-time QRWs to build
potential graph invariants [11,12].
Berry et al. studied discrete-time quantum walks on the line and on general undirected graphs with two interacting or
noninteracting particles [13].
For strongly regular graphs, they showed that noninteracting discrete-time quantum walks can distinguish some but not all nonisomorphic graphs
with the same family parameters.
Gamble et al. extended
these results, proving that QRWs of two noninteracting particles
will always fail to distinguish pairs of nonisomorphic SRGs with
the same family parameters [14]. Then Rudinger et al.
numerically demonstrated that three-particle noninteracting walks
have distinguishing power on pairs of SRGs [15,16].
In our previous paper [17] we investigated GI problem in strongly regular (SRG) graphs by using the entanglement entropy. We obtained the adjacency matrix of SRG in the stratification basis, then we calculated the entanglement entropy in non-isomorph SRGs and showed that the entanglement entropy can distinguish the non-isomorph pairs of SRGs.

In this paper we use quantum walk to distinguish non-isomorph cospectral graphs. Cospectral graphs are graphs that share the same graph spectrum. The non-isomorph cospectral scalable pairs $G_4(n,n+2)$ and $G_5(n,n+2)$ are introduced in [18]. We use $n$-particle quantum walk for GI problem in these graphs. to this aim we rewrite the adjacency matrices of these two graphs in the new basis. The new basis are obtained by fermionization of $n$-particle standard basis. Then the adjacency matrices of two non-isomorph pairs are different in these fermionic basis. Therefore the $n$-particle quantum walk will be able to distinguish these non-isomorph pairs.

Also we use the adjacency matrices of $G_4$ and $G_5$ to construct two new graphs which we call $T_4(a,b)$ and $T_5(a,b)$. These two graphs are Cospectral and non-isomorph for any positive integer $a$ when $b=a+2$. We use the antisymmetric fermionic basis again and rewrite the adjacency matrices of two new graphs in these basis. From the difference between the new adjacency matrices of two graphs, one can conclude that they are non-isomorph.

Then we discriminate pairs of non-isomorph cospectral graphs by using the entanglement entropy. One of the operational entanglement criteria is the Schmidt decomposition [19-21]. The Schmidt decomposition is a very good
tool to study entanglement of bipartite pure states. The entanglement of a partly entangled pure state can be
naturally parametrized by its entropy of entanglement, defined as
the von Neumann entropy, or equivalently as the Shannon entropy of the squares of the Schmidt
coefficients [19,21]. In our model the nodes of networks are considered as identical quantum oscillators [22]. The ground state wave function is obtained in terms of the Laplacian $L$ which is related to the adjacency matrix of network. Two non-isomorph pairs have the same energy but their ground state wave functions are different. For calculating the entanglement entropy in graphs, we use a method in three stage [23] and use the generalized Schur complement method [23,24].

The paper is structured as follows. In Section 2 we give some preliminaries in four subsections. First we explain some interpretation about the graph and the stratification techniques in 2.1. Then in 2.2 we briefly clarify quantum walk. The model and the Hamiltonian which we used, are described in 2.3. The section 2.4 is about Schmidt decomposition and the entanglement entropy.
In section 3, first we introduce two non-isomorph graphs $G_4(a,b)$ and $G_5(a,b)$ and prove that they are cospectral. Then in 3.1 we investigate GI in these two graphs by using quantum walk. To this aim we introduce antisymmetric fermionic basis and rewrite the adjacency matrices of graphs in these basis. The results show that the $n$-particle quantum walk can distinguish  pairs of non-isomorph graphs $G_4(a,b)$ and $G_5(a,b)$. In 3.1.1 we do the same process for two new non-isomorph graphs $T_4(a,b)$ and $T_5(a,b)$. Then in 3.2 we use the entanglement entropy for distinguishing the non-isomorph pairs: $G_4(a,b)$ and $G_5(a,b)$.
In section 4 we give some examples of non-isomorph cospectral graphs which are distinguished by using single particle quantum walk.
Finally in section 5 we give some other examples that the single particle quantum walk can not distinguish non-isomorphic graphs. The entanglement entropy is used to distinguish these graphs.
We discuss our conclusions in Section 5.
For calculating the entanglement entropy between two arbitrary parts of graphs, we use a method which is described in Appendix A. In Appendix B, we explain the generalized Schur complement method which is used in this paper.

\section{Preliminaries}
\subsection{Graphs and their Stratification techniques}
A graph is a pair $G=(V,E)$, where $V$ is a non-empty set and $E$ is a subset of $\{(i,j);i,j\in
V, i\neq j\}$. Elements of $V$ and of $E$ are called vertices and edges, respectively. Two vertices
$i, j \in V$ are called adjacent if $(i, j ) \in E$, and in that case we write $i\sim j $. A finite sequence
$i_0; i_1; ...; i_n \in V$ is called a walk of length $n$ (or of $n$ steps) if $i_{k-1}\sim i_k$
for all $k = 1, 2, ..., n$. A graph is called connected if any pair of distinct vertices is connected by a walk. The degree or
valency of a vertex $x \in V$ is defined by $\kappa(x) = |{y \in V : y \sim x}|$. The graph structure is fully
represented by the adjacency matrix A defined by
\begin{equation}\label{adj.}
\bigl(A)_{i,j}\;=\;\cases{1 & if $\;i\sim j$
\cr 0 & \mbox{otherwise}\cr}.
\end{equation}
Obviously, (i) $A$ is symmetric; (ii) an element of $A$ takes a value in ${0, 1}$; (iii) a diagonal
element of $A$ vanishes. Let $l_2(V)$ denote the Hilbert space of square-summable functions on
$V$ , and ${|i\rangle; i \in V }$ becomes a complete orthonormal basis of $l_2(V)$. The adjacency matrix is
considered as an operator acting in $l_2(V)$ in such a way that
\begin{equation}
A|i\rangle=\sum_{j\sim i}|j\rangle  \quad\quad i\in V.
\end{equation}
For $i \neq j$ let $\partial(i, j )$ be the length of the shortest walk connecting $i$ and $j$. By definition
$\partial(i, j ) = 0$ for all $i \in V $. The graph becomes a metric space with the distance function $\partial$.
Note that $\partial(i, j ) = 1$ if and only if $i \sim j $. We fix a point $o \in V$ as an origin of the graph.
Then, a natural stratification for the graph is introduced as:
\begin{equation}
V=\bigcup_{i=0}^{\infty}V_i(o)\quad \quad V_i(o):=\{j \in V: \partial (o,j) =i \}
\end{equation}
If $V_k(o) = Ø$ happens for some $k \geq 1$, then $V_l(o) = Ø$ for all $l \geq k$. With each stratum $V_i$, we
associate a unit vector in $l_2(V )$ defined by
\begin{equation}\label{fi}
|\phi_i\rangle=\frac{1}{\sqrt{\kappa_i}}\sum_{k\in V_i(o)}|k\rangle
\end{equation}
where, $\kappa_i:=|V_i(o)|$ and $|k\rangle$ denotes the eigenket of $k$-th vertex at the stratum $i$. The closed
subspace of $l_2(V )$ spanned by ${|\phi_i\rangle}$ is denoted by $\Gamma(G)$. Since ${|\phi_i}$ becomes a complete
orthonormal basis of $\Gamma(G)$, we often write
\begin{equation}
\Gamma(G)=\sum_k \oplus C |\phi_k\rangle
\end{equation}
In this stratification for any connected graph $G$, we have
\begin{equation}
V_1 (\beta)\subseteq V_{i-1}(\alpha)\bigcup V_i(\alpha) \bigcup V_{i+1}(\alpha)
\end{equation}
for each $\beta \in V_i(\alpha)$. Now, recall that the $i$-th adjacency matrix of a graph $G=(V,E)$ is defined
as
\begin{equation}\label{adj.}
\bigl(A_i)_{\alpha,\beta}\;=\;\cases{1 & if $\partial(\alpha,\beta)=i$
\cr 0 & \mbox{otherwise}\cr}.
\end{equation}
Then, for reference state $|\phi_0\rangle$ ($|\phi_0 \rangle= |o\rangle$), with $o \in V$ as reference vertex), we have
\begin{equation}\label{adfi}
A_i|\phi_0\rangle =\sum_{\beta\in V_i(o)}|\beta\rangle.
\end{equation}
Then by using (\ref{fi}) and (\ref{adfi}), we have
\begin{equation}
A_i|\phi_0\rangle =\sqrt{\kappa_i}|\phi_i\rangle.
\end{equation}
For more details you can see [25-27].

\subsection{Continuous time quantum walk}
The continuous-time quantum walk is defined by replacing
Kolmogorov's equation with Schrodinger's equation. Let
$|\phi_{i}(t)\rangle$ be a time-dependent amplitude of the quantum
process on graph $\Gamma$. The wave evolution of the quantum walk
is
\begin{equation}
i\hbar\frac{d}{dt}|\phi(t)\rangle=H|\phi(t)\rangle
\end{equation}
where we assume $\hbar=1$ and $|\phi_{0}\rangle$ is the initial
amplitude wave function of the particle. The solution is given by
\begin{equation}
|\phi_{0}(t)\rangle=e^{-iHt}|\phi_{0}\rangle
\end{equation}

Where elements of
amplitudes between strata are calculated
\begin{equation}
\langle\phi_{i}(t)|\phi_{0}(t)\rangle=\langle\phi_{i}(t)|e^{-iHt}|\phi_{0}\rangle
\end{equation}
Obviously the above result indicates that the amplitudes of
observing walk on vertices belonging to a given stratum are the
same. Actually one can straightforwardly  the transition
probabilities between the vertices depend only on the distance
between the vertices irrespective of which site the walk has
started. So, if stratification of two non-isomorphism graph is
different, the quantum walk on these graphs are different.

\subsection{The model and hamiltonian}
The nodes are considered as identical quantum oscillators, interacting as
dictated by the network topology encoded in the Laplacian $L$. The
Laplacian of a network is defined from the Adjacency matrix as
$L_{ij} = k_i\delta_{ij}- A_{ij}$ , where $k_i =\sum_j A_{ij}$ is
the connectivity of node $i$, i.e., the number of nodes connected
to $i$. The Hamiltonian of the quantum network thus reads:
\begin{equation}\label{Ham}
H=\frac{1}{2}(P^T P+ X^T(I+2gL)X)
\end{equation}
here $I$ is the $N \times N$ identity matrix, $g$ is the coupling
strength between connected oscillators while $p^T=(p_1,p_2,...,
p_N)$ and $x^T=(x_1,x_2, ..., x_N)$ are the operators
corresponding to the momenta and positions of nodes respectively,
satisfying the usual commutation relations: $[x, p^T] = i\hbar I$
(we set $\hbar = 1$ in the following) and the matrix $V=I+2gL$ is
the potential matrix. Then the ground state of this Hamiltonian
is:
\begin{equation}\label{gs}
\psi(X)=\frac{(det(I+2gL))^{1/4}}{\pi^{N/4}}exp(-\frac{1}{2}(X^T(I+2gL)X))
\end{equation}
where the $A_g=\frac{(det(I+2gL))^{1/4}}{\pi^{N/4}}$ is the
normalization factor for wave function. The elements of the
potential matrix in terms of entries of adjacency matrix is
$$V_{ij}=(1+2g\kappa_i)\delta_{ij}-2gA_{ij}$$
The ground state energy is in terms of the eigenvalues of potential matrix,
\begin{equation}\label{gsEnergy}
E_G=\frac{1}{2}\prod_{i=1}^{N}(1+2g\alpha_i)
\end{equation}
where $\alpha_i$s are the eigenvalues of Laplacian matrix, which are written in terms of eigenvalues of adjacency matrix.

The eigenvalues of adjacency matrix in cospectral graphs are the same, so the non-isomorph cospectral graphs have the same ground state energy.
But they have different adjacency matrices, so their ground state wave functions are different. Therefore the entanglement entropy of ground state wave function can distinguish non-isomorph graphs.

\subsection{Schmidt decomposition and entanglement entropy}
The Schmidt decomposition is a very good tool to study
entanglement of bipartite pure states. The Schmidt number provides
an important variable to classify entanglement. Any bipartite pure
state $|\psi\rangle_{AB} \in \textsl{H}=\textsl{H}_A
\otimes\textsl{H}_B$ can be decomposed, by choosing an appropriate
basis, as
\begin{equation}\label{sch}
|\psi\rangle_{AB}=\sum_{i=1}^m
\alpha_i|a_i\rangle\otimes|b_i\rangle
\end{equation}
where $1 \leq m \leq min\{dim(\textsl{H}_A); dim(\textsl{H}_B)\}$,
and $\alpha_i
> 0$ with $\sum_{i=1}^m \alpha_i^2 = 1$. Here $|a_i\rangle$ ($|b_i\rangle$) form a part of an
orthonormal basis in $\textsl{H}_A$ ($\textsl{H}_B$). The positive
numbers $\alpha_i$ are called the Schmidt coefficients of
$|\psi\rangle_{AB}$ and the number $m$ is called the Schmidt rank
of $|\psi\rangle_{AB}$. The entanglement of a partly entangled
pure state can be naturally parameterized by its entropy of
entanglement, defined as the Von Neumann entropy of either
$\rho_A$ or $\rho_B$, or equivalently as the Shannon entropy of
the squares of the Schmidt coefficients [19,21].
\begin{equation}\label{ent}
E=-Tr\rho_A log_2\rho_A= Tr\rho_B log_2\rho_B=-\sum_i\alpha_i^2
log_2 \alpha_i^2
\end{equation}

\section{Investigation of graph isomorphism (GI) problem in $G_4(a,b)$ and $G_5(a,b)$}
In this section, the graphs $G_4(a,b)$ and $G_5(a,b)$ with $2a + 6b$
vertices are defined. The $(n + 2)$-regular graphs $G_4(n,n + 2)$
and $G_5(n,n + 2)$ are a pair of connected cospectral integral
regular graphs for any positive integer n. We prove that these two
graphs are non isomorphic by using the entanglement entropy.
The adjacency of $G_4(a,b)$ are defined as
\begin{equation}
A(G_4(a,b))=\left(\begin{array}{cc}
          A_0& A_1\\
            A_1& A_0\\
          \end{array}\right)
\end{equation}
where
\begin{equation}\label{A0G4}
A_0(G_4)=\left(\begin{array}{cccc}
          0& J_{ab} & 0 & 0\\
          J_{ba}& 0 & I_b & 0\\
          0 & I_b & 0 & B_b\\
          0 & 0 & B_b & 0\\
          \end{array}\right)
\end{equation}
and
\begin{equation}\label{A1G4}
A_1(G_4)=\left(\begin{array}{cccc}
          0& 0 & 0 & 0\\
          0& I_b & 0 & 0\\
          0 & 0 & 0 & 0\\
          0 & 0 & 0 & I_b\\
          \end{array}\right)
\end{equation}
and
\begin{equation}\label{B45}
B=\left(\begin{array}{ccc}
          1& J_{1,(b-2)} & 0\\
          J_{(b-2),1}& J_{b-2}-I_{b-2} & J_{(b-2),1}\\
          0 & J_{1,(b-2)} & 1 \\
          \end{array}\right)
\end{equation}

After some relabeling,  the total adjacency matrix for $G_4(a,b)$ is
\begin{equation}
A(G_4(a,b))=\left(\begin{array}{cccccccc}
          0& 0 & J_{ab} & 0 & 0 & 0 & 0 & 0\\
          0& 0 & I_{b} & B_b & 0 & 0 & 0 & 0\\
          J_{ba} & I_b & 0 & 0 & I_b & 0 & 0 & 0\\
          0 & B_b & 0 & 0 & 0 & I_b & 0 & 0\\
          0 & 0 & I_b & 0 & 0 & 0 & J_{ba} & I_b \\
          0 & 0 & 0 & I_b & 0 & 0 & 0 & B_b\\
          0 & 0 & 0 & 0 & J_{ab} & 0 & 0 & 0\\
          0 & 0 & 0 & 0 & I_b & B_b & 0 & 0\\
          \end{array}\right)
\end{equation}
The adjacency matrix for $G_5(a,b)$ is

\begin{equation}
A(G_5(a,b))=\left(\begin{array}{cc}
          A_0& A_1\\
            A_1& A_0\\
          \end{array}\right)
\end{equation}

where $A_0$ and $A_1$  for $G_5(a,b)$ are
\begin{equation}\label{A0G5}
A_0(G_5)=\left(\begin{array}{cccc}
          0& J_{ab} & 0 & 0\\
          J_{ba}& 0 & I_b & I_b\\
          0 & I_b & 0 & 0\\
          0 & I_b & 0 & 0\\
          \end{array}\right)
\end{equation}
and
\begin{equation}\label{A1G5}
A_1(G_5)=\left(\begin{array}{cccc}
          0& 0 & 0 & 0\\
          0& 0 & 0 & 0\\
          0 & 0 & B_b & 0\\
          0 & 0 & 0 & B_b\\
          \end{array}\right)
\end{equation}
and the matrix $B$ is the same as the $G_4(a,b)$.
After some relabeling, the adjacency matrix of $G_5(a,b)$ is
\begin{equation}
A(G_5(a,b))=\left(\begin{array}{cccccccc}
          0& J_{ab} & 0 & 0 & 0 & 0 & 0 & 0\\
          J_{ba}& 0 & I_{b} & I_b & 0 & 0 & 0 & 0\\
          0 & I_b & 0 & 0 & B_b & 0 & 0 & 0\\
          0 & I_b & 0 & 0 & 0 & B_b & 0 & 0\\
          0 & 0 & B_b & 0 & 0 & 0 & I_b & 0 \\
          0 & 0 & 0 & B_b & 0 & 0 & I_b & 0\\
          0 & 0 & 0 & 0 & I_b & I_b & 0 & J_{ba}\\
          0 & 0 & 0 & 0 & 0 & 0 & J_{ab} & 0\\
          \end{array}\right)
\end{equation}

\setlength{\unitlength}{0.75cm}
\begin{picture}(6,14)
\linethickness{0.075mm}

\put(2.2,0.8){$10$}
\put(2.2,1.8){$9$}
\put(2.2,2.8){$8$}
\put(2.2,4.8){$7$}
\put(2.2,5.8){$6$}
\put(2.2,6.8){$5$}
\put(2.2,8.8){$4$}
\put(2.2,9.8){$3$}
\put(2.2,10.8){$2$}
\put(2.2,12.8){$1$}

\put(2,1){\circle*{0.2}}
\put(2,2){\circle*{0.2}}
\put(2,3){\circle*{0.2}}
\put(2,5){\circle*{0.2}}
\put(2,6){\circle*{0.2}}
\put(2,7){\circle*{0.2}}
\put(2,9){\circle*{0.2}}
\put(2,10){\circle*{0.2}}
\put(2,11){\circle*{0.2}}
\put(2,13){\circle*{0.2}}

\put(5.3,0.8){$19$}
\put(5.3,1.8){$18$}
\put(5.3,2.8){$17$}
\put(5.3,4.8){$16$}
\put(5.3,5.8){$15$}
\put(5.3,6.8){$14$}
\put(5.3,8.8){$13$}
\put(5.3,9.8){$12$}
\put(5.3,10.8){$11$}
\put(5.3,12.8){$20$}

\put(6,1){\circle*{0.2}}
\put(6,2){\circle*{0.2}}
\put(6,3){\circle*{0.2}}
\put(6,5){\circle*{0.2}}
\put(6,6){\circle*{0.2}}
\put(6,7){\circle*{0.2}}
\put(6,9){\circle*{0.2}}
\put(6,10){\circle*{0.2}}
\put(6,11){\circle*{0.2}}
\put(6,13){\circle*{0.2}}

\put(2,1){\line(1,0){4}}
\put(2,2){\line(1,0){4}}
\put(2,3){\line(1,0){4}}
\put(2,9){\line(1,0){4}}
\put(2,10){\line(1,0){4}}
\put(2,11){\line(1,0){4}}

\put(2,13){\circle{2}}
\put(6,13){\circle{2}}
\put(2,10){\oval(1.8,3)}
\put(2,6){\oval(1.8,3)}
\put(2,2){\oval(1.8,3)}
\put(6,10){\oval(1.8,3)}
\put(6,6){\oval(1.8,3)}
\put(6,2){\oval(1.8,3)}

\put(2,12){\oval(1,2)[l]}
\put(2,11.5){\oval(1.5,3)[l]}
\put(2,11){\oval(2,4)[l]}
\put(2,9){\oval(0.5,4)[l]}
\put(2,8){\oval(1,4)[l]}
\put(2,7){\oval(1.5,4)[l]}

\put(2,5){\oval(0.5,4)[l]}
\put(2,4.5){\oval(2,5)[l]}
\put(2,4.5){\oval(2.5,3)[l]}
\put(2,3.5){\oval(3,5)[l]}
\put(2,3.5){\oval(1,3)[l]}
\put(2,3){\oval(4,4)[l]}

\put(0.5,12){$J$}
\put(0.5,8){$I$}
\put(4,11.5){$I$}
\put(4,3.5){$I$}
\put(-0.5,4){$B$}
\put(7.5,12){$J$}
\put(7.5,8){$I$}
\put(8,4){$B$}

\put(6,12){\oval(1,2)[r]}
\put(6,11.5){\oval(1.5,3)[r]}
\put(6,11){\oval(2,4)[r]}
\put(6,9){\oval(0.5,4)[r]}
\put(6,8){\oval(1,4)[r]}
\put(6,7){\oval(1.5,4)[r]}

\put(6,5){\oval(0.5,4)[r]}
\put(6,4.5){\oval(2,5)[r]}
\put(6,4.5){\oval(2.5,3)[r]}
\put(6,3.5){\oval(3,5)[r]}
\put(6,3.5){\oval(1,3)[r]}
\put(6,3){\oval(4,4)[r]}

\put(12.2,0.8){$10$}
\put(12.2,1.8){$9$}
\put(12.2,2.8){$8$}
\put(12.2,4.8){$7$}
\put(12.2,5.8){$6$}
\put(12.2,6.8){$5$}
\put(12.2,8.8){$4$}
\put(12.2,9.8){$3$}
\put(12.2,10.8){$2$}
\put(12.2,12.8){$1$}

\put(12,1){\circle*{0.2}}
\put(12,2){\circle*{0.2}}
\put(12,3){\circle*{0.2}}
\put(12,5){\circle*{0.2}}
\put(12,6){\circle*{0.2}}
\put(12,7){\circle*{0.2}}
\put(12,9){\circle*{0.2}}
\put(12,10){\circle*{0.2}}
\put(12,11){\circle*{0.2}}
\put(12,13){\circle*{0.2}}

\put(15.3,0.8){$19$}
\put(15.3,1.8){$18$}
\put(15.3,2.8){$17$}
\put(15.3,4.8){$16$}
\put(15.3,5.8){$15$}
\put(15.3,6.8){$14$}
\put(15.3,8.8){$13$}
\put(15.3,9.8){$12$}
\put(15.3,10.8){$11$}
\put(15.3,12.8){$20$}

\put(16,1){\circle*{0.2}}
\put(16,2){\circle*{0.2}}
\put(16,3){\circle*{0.2}}
\put(16,5){\circle*{0.2}}
\put(16,6){\circle*{0.2}}
\put(16,7){\circle*{0.2}}
\put(16,9){\circle*{0.2}}
\put(16,10){\circle*{0.2}}
\put(16,11){\circle*{0.2}}
\put(16,13){\circle*{0.2}}

\put(12,7){\line(1,0){4}}
\put(12,5){\line(1,0){4}}
\put(12,3){\line(1,0){4}}
\put(12,1){\line(1,0){4}}
\put(12,7){\line(4,-1){4}}
\put(12,6){\line(4,1){4}}
\put(12,6){\line(4,-1){4}}
\put(12,5){\line(4,1){4}}
\put(12,3){\line(4,-1){4}}
\put(12,2){\line(4,1){4}}
\put(12,2){\line(4,-1){4}}
\put(12,1){\line(4,1){4}}

\put(12,13){\circle{2}}
\put(16,13){\circle{2}}
\put(12,10){\oval(1.8,3)}
\put(12,6){\oval(1.8,3)}
\put(12,2){\oval(1.8,3)}
\put(16,10){\oval(1.8,3)}
\put(16,6){\oval(1.8,3)}
\put(16,2){\oval(1.8,3)}

\put(12,12){\oval(1,2)[l]}
\put(12,11.5){\oval(1.5,3)[l]}
\put(12,11){\oval(2,4)[l]}
\put(12,9){\oval(0.5,4)[l]}
\put(12,8){\oval(1,4)[l]}
\put(12,7){\oval(1.5,4)[l]}

\put(12,7){\oval(2.5,8)[l]}
\put(12,6){\oval(3,8)[l]}
\put(12,5){\oval(3.5,8)[l]}

\put(10.5,12){$J$}
\put(10,9){$I$}
\put(9.5,5){$I$}
\put(18.5,5){$I$}
\put(14,7.5){$B$}
\put(17.5,12){$J$}
\put(18,9){$I$}
\put(14,3.5){$B$}

\put(16,12){\oval(1,2)[r]}
\put(16,11.5){\oval(1.5,3)[r]}
\put(16,11){\oval(2,4)[r]}
\put(16,9){\oval(0.5,4)[r]}
\put(16,8){\oval(1,4)[r]}
\put(16,7){\oval(1.5,4)[r]}

\put(16,7){\oval(2.5,8)[r]}
\put(16,6){\oval(3,8)[r]}
\put(16,5){\oval(3.5,8)[r]}

\put(4,0){$(1)$}
\put(14,0){$(2)$}
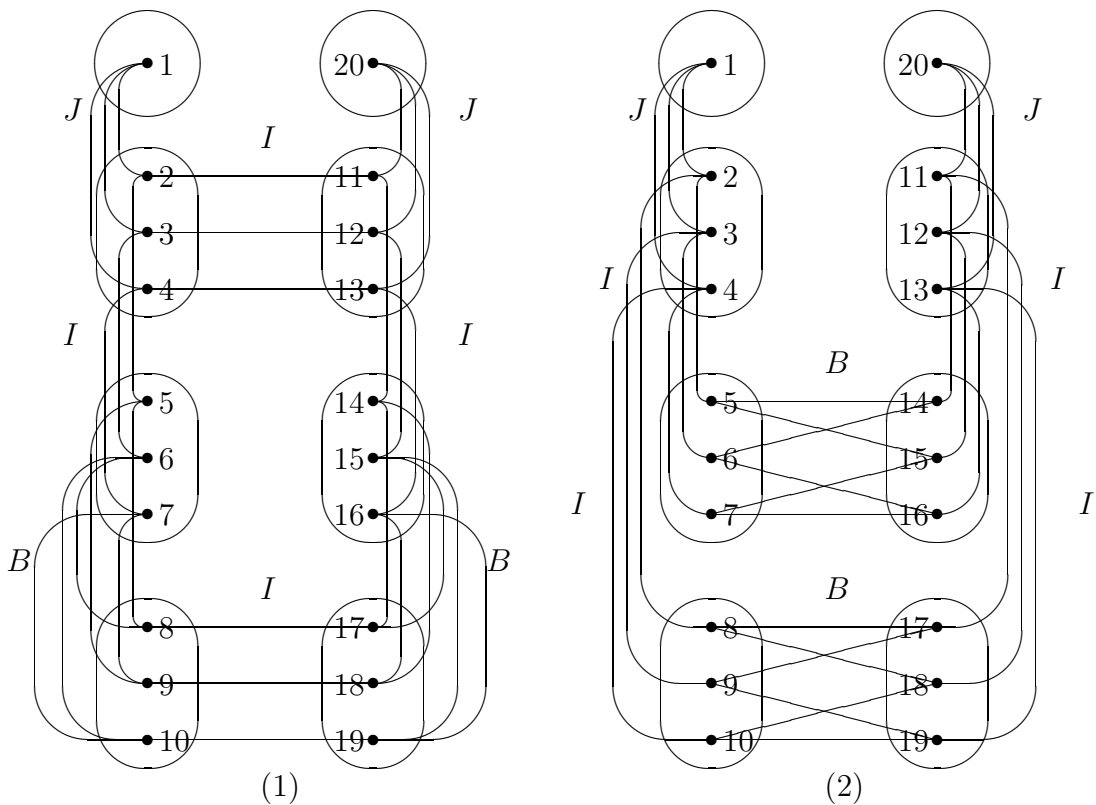
\put(2,-1){\footnotesize FIG I: An example of $G_4(a,b)$ in $(1)$ and $G_5(a,b)$ in $(2)$ with $a=1$ and $b=3$.}
\end{picture}
\newline
\newline
\newline
Now we want to show that two graphs $G_4(a,b)$ and $G_5(a,b)$ are cospectral.
The adjacency matrices of these graphs can be written as
$$A=I_2\otimes A_0+ \sigma_x \otimes A_1$$
So the eigenvalues of adjacency matrices of these two graphs will be the eigenvalues of two matrices $A_0\pm A_1$.
\begin{equation}
(A_0\pm A_1)(G_4)=\left(\begin{array}{cccc}
          0& J_{ab} & 0 & 0\\
          J_{ba}& \pm I_b & I_b & 0\\
          0 & I_b & 0 & B_b\\
          0 & 0 & B_b & \pm I_b\\
          \end{array}\right)
\end{equation}
We want to diagonalize the blocks of above matrix. So we can apply following transformation
\begin{equation}
          =\left(\begin{array}{cccc}
          O_1^T& 0 & 0 & 0\\
            0& O_2^T & 0 & 0 \\
           0 & 0 & O_3^T & 0\\
           0 & 0 & 0 & O_4^T\\
          \end{array}\right)\left(\begin{array}{cccc}
          0& J_{ab} & 0 & 0\\
          J_{ba}& \pm I_b & I_b & 0\\
          0 & I_b & 0 & B_b\\
          0 & 0 & B_b & \pm I_b\\
          \end{array}\right)\left(\begin{array}{cccc}
          O_1& 0 & 0 & 0\\
            0& O_2 & 0 & 0 \\
           0 & 0 & O_3 & 0\\
           0 & 0 & 0 & O_4\\
          \end{array}\right)
\end{equation}
$$=\left(\begin{array}{cccc}
          0& O_1^T J_{ab}O_2 & 0 & 0\\
          O_2^TJ_{ba}O_1& \pm O_2^TO_2 & O_2^TO_3 & 0\\
          0 & O_3^TO_2 & 0 & O_3^TB_bO_4\\
          0 & 0 & O_4^TB_bO_3 & \pm O_4^TO_4\\
          \end{array}\right)$$
Then by choosing $O_2=O_3=O_4$, the transformed matrix will be
\begin{equation}
\left(\begin{array}{cccc}
          0& SVD(J_{ab}) & 0 & 0\\
          SVD(J_{ba})& \pm I_b & I_b & 0\\
          0 & I_b & 0 & D_B\\
          0 & 0 & D_B & \pm I_b\\
          \end{array}\right)
\end{equation}
Therefore the eigenvalues of $G_4$ will be the eigenvalues of these matrices:
\begin{equation}
\left(\begin{array}{cccc}
          0& \sqrt{ab} & 0 & 0\\
          \sqrt{ab}& \pm 1 & 1 & 0\\
          0 & 1 & 0 & b-1\\
          0 & 0 & b-1 & \pm 1\\
          \end{array}\right),\quad\quad\quad\left(\begin{array}{ccc}
          \pm1 & 1 & 0\\
          1& 0 & 1\\
          0 & 1 & \pm 1\\
          \end{array}\right)
\end{equation}
By cosidering $a=n$ and $b=n+2$, the eigenvalues will be
$$\pm(n+2), \underbrace{\pm(n+1)}_{2 times}, \pm n, \underbrace{\pm 2}_{(b-1) times}, \underbrace{\pm 1}_{2(b-1) times}$$

The same process can be applied to graph $G_5$, So the eigenvalues of adjacency matrix of graph $G_5$ will be the eigenvalues of these matrices:
\begin{equation}
\left(\begin{array}{cccc}
          0& \sqrt{ab} & 0 & 0\\
          \sqrt{ab}& 0 & 1 & 1\\
          0 & 1 & \pm(b-1) & 0\\
          0 & 1 & 0 & \pm(b-1)\\
          \end{array}\right),\quad\quad\quad\left(\begin{array}{ccc}
          0 & 1 & 1\\
          1& \pm 1 & 0\\
          1 & 0 & \pm 1\\
          \end{array}\right)
\end{equation}
Again by considering $a=n$ and $b=n+2$, the eigenvalues will be
$$\pm(n+2),\underbrace{\pm(n+1)}_{2 times},\pm n, \underbrace{\pm 2}_{(b-1) times}, \underbrace{\pm 1}_{2(b-1) times}$$
So these two graphs for all $b=a+2$ are cospectral.

\subsection{Investigation of GI problem via quantum walk in the antisymmetric fermionic basis}
Now we want to use quantum walk for investigating graph isomorphism problem in these two graphs.
The total adjacency matrix for $G_4(a,b)$ can be written as
\begin{equation}
A(G_4(a,b))=\left(\begin{array}{cccccccc}
          0& J_{ab} & 0 & 0 & 0 & 0 & 0 & 0\\
          J_{ba} & 0 & I_{b} & 0 & I_{b} & 0 & 0 & 0\\
          0 & I_b & 0 & B_{b} & 0 & 0 & 0 & 0\\
          0 & 0 & B_{b} & 0 & 0 & 0 & I_{b} & 0\\
          0 & I_{b} & 0 & 0 & 0 & I_{b} & 0 & J_{ba} \\
          0 & 0 & 0 & 0 & I_{b} & 0 & B_b & 0 \\
          0 & 0 & 0 & I_{b} & 0 & B_{b} & 0 & 0\\
          0 & 0 & 0 & 0 & J_{ab} & 0 & 0 & 0\\
          \end{array}\right)
\end{equation}
And the adjacency matrix of $G_5(a,b)$ can be written as

\begin{equation}
A(G_5(a,b))=\left(\begin{array}{cccccccc}
          0& J_{ab} & 0 & 0 & 0 & 0 & 0 & 0\\
          J_{ba}& 0 & I_{b} & I_b & 0 & 0 & 0 & 0\\
          0 & I_b & 0 & 0 & 0 & B_{b} & 0 & 0\\
          0 & I_b & 0 & 0 & 0 & 0 & B_{b} & 0\\
          0 & 0 & 0 & 0 & 0 & I_{b} & I_b & J_{ba} \\
          0 & 0 & B_{b} & 0 & I_{b} & 0 & 0 & 0\\
          0 & 0 & 0 & B_{b} & I_b & 0 & 0 & 0\\
          0 & 0 & 0 & 0 & J_{ab} & 0 & 0 & 0\\
          \end{array}\right)
\end{equation}

We want to rewrite the adjacency matrices of these two graphs in the new basis.

The strata of $G_4(a,b)$ and $G_5(a,b)$ are obtained by
fermionization as following form
$$|\phi_{0}\rangle=\frac{1}{\sqrt{a!}}\sum_{i_{1},i_{2},...,i_{a}}\varepsilon_{i_{1},i_{2},...,i_{a}}|i_{1}\rangle\otimes|i_{2}\rangle\otimes...\otimes|i_{a}\rangle$$
$$|\phi_{l}\rangle=\frac{1}{\sqrt{a!}\sqrt{ab}}\sum_{i_{1},i_{2},...,i_{a}}\varepsilon_{i_{1},i_{2},...,i_{a}}|i_{1}\rangle\otimes|i_{2}\rangle\otimes...\otimes|i_{k-1}\rangle
(\sum_{j=1}^{b}|a+(l-1)b+j\rangle)\otimes|i_{k+1}\rangle...\otimes|i_{a}\rangle,\quad\quad(l=1,...,6)$$
\begin{equation}\label{fermiBasis}
|\phi_{7}\rangle=\frac{1}{\sqrt{a!}}\sum_{i_{1},i_{2},...,i_{a}}\varepsilon_{i_{1},i_{2},...,i_{a}1,2,...,a}|a+6b+i_{1}\rangle\otimes|a+6b+i_{2}\rangle\otimes...\otimes|a+6b+i_{a}\rangle
\end{equation}
The dimension of this fermionic space is
$\left(\begin{array}{c}
  n \\
 a \\
\end{array}
\right)$
But we choose the above antisymmetric $a$-particle fermionic basis for graphs $G_4(a,b)$ and $G_5(a,b)$.
We want to apply the following adjacency matrices of two graphs on the defined basis.
\begin{equation}
A=\sum_{i}I\otimes I\otimes...\otimes \underbrace{A_1}_{i}\otimes
I...\otimes I
\end{equation}
where $I$ is identity matrix.
Now, by applying adjaceny matrix of $G_4(a,b)$ and $G_5(a,b)$ on
the new basis, we have
$$A_{G_4(a,b)}|\phi_{0}\rangle=\sqrt{ab}|\phi_{1}\rangle$$
$$A_{G_4(a,b)}|\phi_{1}\rangle=\sqrt{ab}|\phi_{0}\rangle+|\phi_{2}\rangle+|\phi_{4}\rangle$$
$$A_{G_4(a,b)}|\phi_{2}\rangle=|\phi_{1}\rangle+(b-1)|\phi_{3}\rangle$$
$$A_{G_4(a,b)}|\phi_{3}\rangle=(b-1)|\phi_{2}\rangle+|\phi_{6}\rangle$$
$$A_{G_4(a,b)}|\phi_{4}\rangle=\sqrt{ab}|\phi_{7}\rangle$$
$$A_{G_4(a,b)}|\phi_{5}\rangle=(b-1)|\phi_{6}\rangle+|\phi_{4}\rangle$$
$$A_{G_4(a,b)}|\phi_{6}\rangle=|\phi_{3}\rangle+(b-1)|\phi_{5}\rangle$$
\begin{equation}
A_{G_4(a,b)}|\phi_{7}\rangle=\sqrt{ab}|\phi_{4}\rangle
\end{equation}
So, the adjacency matrix in the stratification basis is
\begin{equation}
A_{G_4(a,b)}=\left(%
\begin{array}{cccccccc}
  0 & \sqrt{ab} & 0 & 0 & 0 & 0 & 0 & 0 \\
  \sqrt{ab}& 0 & 1 & 0 & 1 & 0 & 0 & 0 \\
  0 & 1 & 0 & b-1 & 0 & 0 & 0 & 0 \\
  0 & 0 & b-1 & 0 & 0 & 0 & 1 & 0 \\
  0 & 1 & 0 & 0 & 0 & 1 & 0 & \sqrt{ab} \\
  0 & 0 & 0 & 0 & 1 & 0 & b-1 & 0 \\
  0 & 0 & 0 & 1 & 0 & b-1 & 0 & 0 \\
  0 & 0 & 0 & 0 & \sqrt{ab} & 0 & 0 & 0 \\
\end{array}%
\right)
\end{equation}
and
$$A_{G_5(a,b)}|\phi_{0}\rangle=\sqrt{ab}|\phi_{1}\rangle$$
$$A_{G_5(a,b)}|\phi_{1}\rangle=\sqrt{ab}|\phi_{0}\rangle+|\phi_{2}\rangle+|\phi_{3}\rangle$$
$$A_{G_5(a,b)}|\phi_{2}\rangle=|\phi_{1}\rangle+(b-1)|\phi_{5}\rangle$$
$$A_{G_5(a,b)}|\phi_{3}\rangle=|\phi_{1}\rangle+(b-1)|\phi_{6}\rangle$$
$$A_{G_5(a,b)}|\phi_{4}\rangle=|\phi_{5}\rangle+|\phi_{6}\rangle+\sqrt{ab}|\phi_{7}\rangle$$
$$A_{G_5(a,b)}|\phi_{5}\rangle=(b-1)|\phi_{2}\rangle+|\phi_{4}\rangle$$
$$A_{G_5(a,b)}|\phi_{6}\rangle=(b-1)|\phi_{3}\rangle+|\phi_{4}\rangle$$
\begin{equation}
A_{G_5(a,b)}|\phi_{7}\rangle=\sqrt{ab}|\phi_{4}\rangle
\end{equation}
So, the adjacency matrix in the stratification basis is
\begin{equation}
A_{G_5(a,b)}=\left(%
\begin{array}{cccccccc}
  0 & \sqrt{ab} & 0 & 0 & 0 & 0 & 0 & 0 \\
  \sqrt{ab}& 0 & 1 & 1 & 0 & 0 & 0 & 0 \\
  0 & 1 & 0 &0 & 0 & b-1 & 0 & 0 \\
  0 & 1 & 0 & 0 & 0 & 0 & b-1 & 0 \\
  0 & 0 & 0 & 0 & 0 & 1 & 1 & \sqrt{ab} \\
  0 & 0 & b-1 & 0 & 1 & 0 & 0 & 0 \\
  0 & 0 & 0 & b-1 & 1 & 0 & 0 & 0 \\
  0 & 0 & 0 & 0 & \sqrt{ab} & 0 & 0 & 0 \\
\end{array}%
\right)
\end{equation}
We see that the adjacency matrices of two above graphs are different. So,
non-isomorphism of two cospectral graph cab be determined by
$n$-particle quantum walk.

\subsubsection{Investigation of GI problem via quantum walk in $T_4(a,b)$ and $T_5(a,b)$}
We can construct two nonisomorph graphs similar to $G_4(a,b)$ and $G_5(a,b)$ by replacing the $A_0$ and $A_1$ in adjacency matrices. The new graphs $T_4(a,b)$ and $T_5(a,b)$ are cospectral and non-isomorph for $b=a+2$.
\begin{equation}
A=\left(\begin{array}{cc}
          A_1& A_0\\
            A_0& A_1\\
          \end{array}\right)
\end{equation}
Where $A_{0},A_{1}$  are the same  as (\ref{A0G4}), (\ref{A1G4}) for $T_4$ and (\ref{A0G5}), (\ref{A1G5}) for $T_5$. We use the antisymmetric fermionic basis of (\ref{fermiBasis}) .Then, by applying adjaceny matrix
of $T_4(a,b)$ and $T_5(a,b)$ on these basis, we have
$$A_{T_4(a,b)}|\phi_{0}\rangle=\sqrt{ab}|\phi_{4}\rangle$$
$$A_{T_4(a,b)}|\phi_{1}\rangle=\sqrt{ab}|\phi_{7}\rangle+|\phi_{1}\rangle+|\phi_{5}\rangle$$
$$A_{T_4(a,b)}|\phi_{2}\rangle=|\phi_{4}\rangle+(b-1)|\phi_{6}\rangle$$
$$A_{T_4(a,b)}|\phi_{3}\rangle=(b-1)|\phi_{5}\rangle+|\phi_{3}\rangle$$
$$A_{T_4(a,b)}|\phi_{4}\rangle=\sqrt{ab}|\phi_{0}\rangle+|\phi_{2}\rangle+|\phi_{4}\rangle$$
$$A_{T_4(a,b)}|\phi_{5}\rangle=(b-1)|\phi_{3}\rangle+|\phi_{1}\rangle$$
$$A_{T_4(a,b)}|\phi_{6}\rangle=|\phi_{6}\rangle+(b-1)|\phi_{2}\rangle$$
\begin{equation}
A_{T_4(a,b)}|\phi_{7}\rangle=\sqrt{ab}|\phi_{1}\rangle
\end{equation}
So, the adjacency matrix of $T_4(a,b)$ in the antisymmetric fermionic basis is
\begin{equation}
A_{T_4(a,b)}=\left(%
\begin{array}{cccccccc}
  0 & 0 & 0 & 0 & \sqrt{ab} & 0 & 0 & 0 \\
  0 & 1 & 0 & 0 & 0 & 1 & 0 & \sqrt{ab} \\
  0 & 0 & 0 & 0 & 1 & 0 & b-1 & 0 \\
  0 & 0 & 0 & 1 & 0 & b-1 & 0 & 0 \\
  \sqrt{ab} & 0 & 1 & 0 & 1 & 0 & 0 & 0 \\
  0 & 1 & 0 & b-1 & 0 & 0 & 0 & 0 \\
  0 & 0 & b-1 & 0 & 0 & 0 & 1 & 0 \\
  0 & \sqrt{ab} & 0 & 0 & 0 & 0 & 0 & 0 \\
\end{array}%
\right)
\end{equation}
And
$$A_{T_5(a,b)}|\phi_{0}\rangle=\sqrt{ab}|\phi_{4}\rangle$$
$$A_{T_5(a,b)}|\phi_{1}\rangle=\sqrt{ab}|\phi_{7}\rangle+|\phi_{5}\rangle+|\phi_{6}\rangle$$
$$A_{T_5(a,b)}|\phi_{2}\rangle=|\phi_{4}\rangle+(b-1)|\phi_{2}\rangle$$
$$A_{T_5(a,b)}|\phi_{3}\rangle=|\phi_{4}\rangle+(b-1)|\phi_{3}\rangle$$
$$A_{T_5(a,b)}|\phi_{4}\rangle=|\phi_{2}\rangle+|\phi_{3}\rangle+\sqrt{ab}|\phi_{0}\rangle$$
$$A_{T_5(a,b)}|\phi_{5}\rangle=(b-1)|\phi_{5}\rangle+|\phi_{1}\rangle$$
$$A_{T_5(a,b)}|\phi_{6}\rangle=(b-1)|\phi_{6}\rangle+|\phi_{1}\rangle$$
\begin{equation}
A_{T_5(a,b)}|\phi_{7}\rangle=\sqrt{ab}|\phi_{1}\rangle
\end{equation}
So, the adjacency matrix of $T_5(a,b)$ in the antisymmetric fermionic basis is
\begin{equation}
A_{T_5(a,b)}=\left(%
\begin{array}{cccccccc}
  0 & 0 & 0 & 0 & \sqrt{ab} & 0 & 0 & 0 \\
  0& 0 & 0 & 0 & 0 & 1 & 1 & \sqrt{ab} \\
  0 & 0 & b-1 &0 & 1 & 0 & 0 & 0 \\
  0 & 0 & 0 & b-1 & 1 & 0 & 0 & 0 \\
  \sqrt{ab} & 0 & 1 & 1 & 0 & 0 & 0 & 0 \\
  0 & 1 & 0 & 0 & 1 & b-1 & 0 & 0 \\
  0 & 1 & 0 & 0 & 0 & 0 & b-1 & 0 \\
  0 & \sqrt{ab} & 0 & 0 & 0 & 0 & 0 & 0 \\
\end{array}%
\right)
\end{equation}
We see that the adjacency matrices of two above graphs are different similar to the cases $G_4(a,b)$ and $G_5(a,b)$. So,
non-isomorphism of two cospectral graph cab be determined by
$n$-particle quantum walk.

\subsection{Investigation of GI by using the entanglement entropy}
In this section we want to use the entanglement entropy to distinguish pairs of non-isomorph graphs. For calculating entanglement entropy between two subsets in a graph, we used a process in three stage [23], which is briefly explained in Appendix A.
The potential matrix $(I+2gL)_{G_4(a,b)}$ will be
$$I+2gL=$$
\begin{equation}
(1+2gb)I_{2a+6b}+\left(\begin{array}{cccccccc}
          0& 0 & -2gJ_{ab} & 0 & 0 & 0 & 0 & 0\\
          0& 0 & -2gI_{b} & -2gB_b & 0 & 0 & 0 & 0\\
          -2gJ_{ba} & -2gI_b & 0 & 0 & -2gI_b & 0 & 0 & 0\\
          0 & -2gB_b & 0 & 0 & 0 & -2gI_b & 0 & 0\\
          0 & 0 & -2gI_b & 0 & 0 & 0 & -2gJ_{ba} & -2gI_b \\
          0 & 0 & 0 & -2gI_b & 0 & 0 & 0 & -2gB_b\\
          0 & 0 & 0 & 0 & -2gJ_{ab} & 0 & 0 & 0\\
          0 & 0 & 0 & 0 & -2gI_b & -2gB_b & 0 & 0\
          \end{array}\right)
\end{equation}

By using our Schur complement method of Appendix B, we
have

$$\widetilde{A}_{22}=A_{22}-A_{12}^T A_{11}^{-1}A_{12}$$
\begin{equation}
\widetilde{A}_{33}=A_{33}-A_{34}A_{44}^{-1}A_{34}^T
\end{equation}

But for $G_4$, $A_{11}$ and $A_{44} \propto I$ So
\begin{equation}
 A_{12}^T A_{12}=A_{34} A_{34}^T=4g^2\left(\begin{array}{cc}
          aJ_b+I_b& B_b\\
          B_b& B_b^2 \\
          \end{array}\right)
\end{equation}

Where $B_b^2=aJ_b+I_b$

So $\widetilde{A}_{22}=\widetilde{A}_{33}=$
\begin{equation}
\left(\begin{array}{cc}
          (1+2gb)I_b-\frac{4g^2}{1+2gb}(aJ_b+I_b)& -\frac{4g^2}{1+2gb}B_b\\
          -\frac{4g^2}{1+2gb}B_b& (1+2gb)I_b-\frac{4g^2}{1+2gb}(aJ_b+I_b) \\
          \end{array}\right)
\end{equation}

and $A_{23}=-2gI_{2b}$.

Therefore we can calculate bipartite entanglement by using three
stages which are introduced in Appendix A. In this case the
potential matrix has a simple form

\begin{equation}
\left(\begin{array}{cc}
          A & B \\
          B^T & C
          \end{array}\right)=\left(\begin{array}{cc}
          \widetilde{A}_{22} & -2gI_{2b} \\
          -2gI_{2b} & \widetilde{A}_{22}
          \end{array}\right)
\end{equation}

So it's clear that the parameter $d_i$s come from
\begin{equation}
\left(\begin{array}{cc}
          I & D_A^{-1} \\
          D_A^{-1} & I
          \end{array}\right)=\left(\begin{array}{cccccccc}
          1 & 0 & \ldots & 0 & 1/\lambda_1 & 0 & \ldots & 0\\
          0 & 1 & \ldots & 0 & 0 & 1/\lambda_2 & \ldots & 0\\
          \vdots & \vdots & \ddots & \vdots & \vdots & \vdots & \ddots & \vdots\\
          0 & 0 & \ldots & 1 & 0 & 0 & \ldots & 1/\lambda_{2b}\\
          1/\lambda_1 & 0 & \ldots & 0 & 1 & 0 & \ldots & 0 \\
          0 & 1/\lambda_2 & \ldots & 0 & 0 & 1 & \ldots & 0 \\
          \vdots & \vdots & \ddots & \vdots & \vdots & \vdots & \ddots & \vdots\\
          0 & 0 & \ldots & 1/\lambda_{2b} & 0 & 0 & \ldots & 1\\
          \end{array}\right)
\end{equation}
Where $\lambda_i$s are the eigenvalues of $\widetilde{A}_{22}$.

The eigenvalues of  matrix $\left(\begin{array}{cc}
          aJ_b+I_b& B_b\\
          B_b& B_b^2 \\
          \end{array}\right)$ are
\begin{equation}
\{ab+b, ab-b+2, \underbrace{2,\ldots,2}_{b-1}
,\underbrace{0,\ldots,0}_{b-1}\}
\end{equation}

So the parameters $d_i$s will be

$$d_1=\frac{2g}{(1+2gb)-\frac{4g^2}{1+2gb}(ab+b)}$$
$$d_2=\frac{2g}{(1+2gb)-\frac{4g^2}{1+2gb}(ab-b+2)}$$
$$d_3=\frac{2g}{(1+2gb)-\frac{8g^2}{1+2gb}}$$
$$\vdots$$
$$d_{b+2}=\frac{2g}{(1+2gb)-\frac{8g^2}{1+2gb}}$$
$$d_{b+3}=\frac{2g}{(1+2gb)}$$
$$\vdots$$
$$d_{2b}=\frac{2g}{(1+2gb)}$$

The potential matrix $(I+2gL)_{G_5(a,b)}$ is
$$I+2gL=$$
\begin{equation}
(1+2gb)I_{2a+6b}+\left(\begin{array}{cccccccc}
          0& -2gJ_{ab} & 0 & 0 & 0 & 0 & 0 & 0\\
          -2gJ_{ba}& 0 & -2gI_{b} & -2gI_b & 0 & 0 & 0 & 0\\
          0 & -2gI_b & 0 & 0 & -2gB_b & 0 & 0 & 0\\
          0 & -2gI_b & 0 & 0 & 0 & -2gB_b & 0 & 0\\
          0 & 0 & -2gB_b & 0 & 0 & 0 & -2gI_b & 0 \\
          0 & 0 & 0 & -2gB_b & 0 & 0 & -2gI_b & 0\\
          0 & 0 & 0 & 0 & -2gI_b & -2gI_b & 0 & -2gJ_{ba}\\
          0 & 0 & 0 & 0 & 0 & 0 & -2gJ_{ab} & 0\\
          \end{array}\right)
\end{equation}

By using our Schur complement method for the first and the last $a
\times a$  parts, the $\widetilde{A_{11}}$ and
$\widetilde{A_{44}}$ will be

\begin{equation}
\widetilde{A_{11}}=\left(\begin{array}{cc}
          (1+2gb)I_a& 0\\
           0 & (1+2gb)I_b-\frac{4g^2a}{(1+2gb)}J_b\\
          \end{array}\right)
\end{equation}

\begin{equation}
\widetilde{A_{44}}=\left(\begin{array}{cc}
          (1+2gb)I_b-\frac{4g^2a}{(1+2gb)}J_b& 0\\
           0 & (1+2gb)I_a\\
          \end{array}\right)
\end{equation}

The inverse of $\widetilde{A_{11}}$ will be

\begin{equation}
\left(\begin{array}{cc}
          \frac{1}{(1+2gb)}I_a & 0 \\
          0& \frac{4g^2a(b-1)-(1+2gb)^2}{4g^2ab(1+2gb)-(1+2gb)^3}I_b+\frac{-4g^2a}{4g^2ab(1+2gb)-(1+2gb)^3}(J_b-I_b) \\
          \end{array}\right)\equiv
          \left(\begin{array}{cc}
          \frac{1}{(1+2gb)}I_a & 0 \\
          0& \alpha I_b +\beta J_b \\
          \end{array}\right)
\end{equation}

So $\widetilde{A_{22}}=A_{22}-A_{12}^T \widetilde{A_{11}}^{-1}
A_{12}$ is in the form

\begin{equation}\label{A22t}
\widetilde{A_{22}}=(1+2gb)\left(\begin{array}{cc}
          I_b& 0\\
           0 & I_b\\
          \end{array}\right)-\left(\begin{array}{cc}
          1& 1\\
           1 & 1\\
          \end{array}\right)\otimes \left(\begin{array}{cc}
          \frac{1}{(1+2gb)}I_a & 0 \\
          0& \alpha I_b+ \beta J_b \\
          \end{array}\right)
\end{equation}

After some similar calculation, we can find the matrix
$\widetilde{A_{33}}=A_{33}-A_{34} \widetilde{A_{44}}^{-1}
A_{34}^T$, same as the matrix $\widetilde{A_{22}}$.

So for calculating the bipartite entanglement between two parts of
the graph $G_5(a,b)$, we have two partite matrix ax
\begin{equation}
\left(\begin{array}{cc}
          A& B\\
           B^T & C\\
          \end{array}\right)=\left(\begin{array}{cc}
          \widetilde{A_{22}}& -2gI_2 \otimes B\\
          -2g I_2 \otimes B & \widetilde{A_{22}}\\
          \end{array}\right)
\end{equation}

It's clear that the matrix $B_b$ in (\ref{B45}) is regular and it
commutes with the matrix $J_b$. So from (\ref{A22t}), we find
that  the  matrices $\widetilde{A_{22}}$ and $I_2 \otimes B$,
commute eachother. Then they can be diagonal simultaneously.
Therefore it is sufficient that we do the stage rescaling in Appendix A. The eigenvalues of $\widetilde{A_{22}}$ are
\begin{equation}
\{\frac{4g^2(ab+2)(1+2gb)-(1+2gb)^3}{4g^2ab-(1+2gb)^2},\underbrace{-\frac{8g^2-(1+2gb)^2}{(1+2gb)}}_{b-1},\underbrace{(1+2gb)}_{b}\}
\end{equation}

The eigenvalues of matrix $-2gI_2 \otimes B$, are
\begin{equation}
\{-2g(b-1),-2g(b-1),-2g,-2g, \underbrace{2g}_{2b-4}\}
\end{equation}

After rescaling, The parameters $d_i$s will be
$$d_1=\frac{-2g(b-1)}{\frac{4g^2(ab+2)(1+2gb)-(1+2gb)^3}{4g^2ab-(1+2gb)^2}}$$
$$d_2=\frac{-2g}{-\frac{8g^2-(1+2gb)^2}{(1+2gb)}}$$
$$d_3=\frac{2g}{-\frac{8g^2-(1+2gb)^2}{(1+2gb)}}$$
$$\vdots$$
$$d_{b}=\frac{2g}{-\frac{8g^2-(1+2gb)^2}{(1+2gb)}}$$
$$d_{b+1}=\frac{-2g(b-1)}{(1+2gb)}$$
$$d_{b+2}=\frac{-2g}{(1+2gb)}$$
$$d_{b+3}=\frac{2g}{(1+2gb)}$$
$$\vdots$$
$$d_{2b}=\frac{2g}{(1+2gb)}$$

\section{Investigation of graph isomorphism via quantum walk in some cospectral graphs}

\textit{Example I}:

Two cospectral nonisomorph graphs $G_1$ and $G_2$ are shown in
Fig (II). They have ten vertices and eighteen edges. The degree distribution of two graphs is $5,5,5,3,3,3,3,3,3,3$.

\setlength{\unitlength}{0.75cm}
\begin{picture}(6,10)
\linethickness{0.075mm}

\put(2,9){$(a)G_1$}
\put(12,9){$(b)G_2$}

\put(2,1.5){$8$}
\put(6,1.5){$9$}
\put(2.8,3.5){$3$}
\put(5,3.5){$4$}
\put(-0.3,5){$7$}
\put(3.9,4.5){$1$}
\put(8.2,5){$10$}
\put(4,6.3){$2$}
\put(2,8.3){$6$}
\put(6,8.3){$5$}

\put(2,2){\circle*{0.2}}
\put(3,4){\circle*{0.2}}
\put(5,4){\circle*{0.2}}
\put(6,2){\circle*{0.2}}
\put(0,5){\circle*{0.2}}
\put(4,5){\circle*{0.2}}
\put(8,5){\circle*{0.2}}
\put(4,6){\circle*{0.2}}
\put(2,8){\circle*{0.2}}
\put(6,8){\circle*{0.2}}

\put(2,2){\line(1,0){4}}
\put(6,2){\line(2,3){2}}
\put(0,5){\line(2,-3){2}}
\put(2,2){\line(3,2){3}}
\put(3,4){\line(1,0){2}}
\put(3,4){\line(3,-2){3}}
\put(0,5){\line(2,3){2}}
\put(0,5){\line(4,1){4}}
\put(3,4){\line(1,2){1}}
\put(3,4){\line(1,1){1}}
\put(4,5){\line(0,1){1}}
\put(4,5){\line(1,-1){1}}
\put(5,4){\line(1,4){1}}
\put(4,6){\line(1,-2){1}}
\put(4,6){\line(4,-1){4}}
\put(2,8){\line(1,0){4}}
\put(2,8){\line(1,-4){1}}
\put(6,8){\line(2,-3){2}}

\put(12,1.5){$8$}
\put(16,1.5){$9$}
\put(12.5,4){$3$}
\put(15.2,4){$4$}
\put(9.7,5){$7$}
\put(13.9,4.5){$1$}
\put(18.2,5){$10$}
\put(14,6.3){$2$}
\put(12,8.3){$6$}
\put(16,8.3){$5$}

\put(12,2){\circle*{0.2}}
\put(13,4){\circle*{0.2}}
\put(15,4){\circle*{0.2}}
\put(16,2){\circle*{0.2}}
\put(10,5){\circle*{0.2}}
\put(14,5){\circle*{0.2}}
\put(18,5){\circle*{0.2}}
\put(14,6){\circle*{0.2}}
\put(12,8){\circle*{0.2}}
\put(16,8){\circle*{0.2}}

\put(12,2){\line(1,0){4}}
\put(16,2){\line(2,3){2}}
\put(10,5){\line(2,-3){2}}
\put(12,2){\line(3,2){3}}
\put(12,2){\line(0,1){6}}
\put(13,4){\line(3,-2){3}}
\put(10,5){\line(2,3){2}}
\put(10,5){\line(4,1){4}}
\put(12,2){\line(2,1){6}}
\put(13,4){\line(1,1){1}}
\put(14,5){\line(0,1){1}}
\put(14,5){\line(1,-1){1}}
\put(15,4){\line(1,4){1}}
\put(12,8){\line(2,-1){6}}
\put(14,6){\line(4,-1){4}}
\put(12,8){\line(1,0){4}}
\put(12,8){\line(1,-4){1}}
\put(16,8){\line(2,-3){2}}

\put(1,0){\footnotesize FIG II: A pair of nonisomorphic cospectral graphs: $(a):G_1$ and $(b):G_2$.}
\put(1,-1){\footnotesize Single particle quantum walk can distinguish these two graphs.}
\end{picture}
\newline
\newline
\newline
The stratification basis are defined in two graph $G_1$ and $G_2$ as
following
$$|\phi_{0}\rangle=|1\rangle$$
$$|\phi_{1}\rangle=\frac{1}{\sqrt{3}}(|2\rangle+|3\rangle+|4\rangle)$$
$$|\phi_{2}\rangle=\frac{1}{\sqrt{3}}(|5\rangle+|7\rangle+|9\rangle)$$
\begin{equation}
|\phi_{3}\rangle=\frac{1}{\sqrt{3}}(|6\rangle+|8\rangle+|10\rangle)
\end{equation}
So
$$A_{G_1}|\phi_{0}\rangle=\sqrt{3}|\phi_{1}\rangle$$
$$A_{G_1}|\phi_{1}\rangle=\sqrt{3}|\phi_{0}\rangle+2|\phi_{1}\rangle+|\phi_{2}\rangle+|\phi_{3}\rangle$$
$$A_{G_1}|\phi_{2}\rangle=|\phi_{1}\rangle+2|\phi_{3}\rangle$$
\begin{equation}
A_{G_1}|\phi_{3}\rangle=|\phi_{1}\rangle+2|\phi_{2}\rangle
\end{equation}
And
$$A_{G_2}|\phi_{0}\rangle=\sqrt{3}|\phi_{1}\rangle$$
$$A_{G_2}|\phi_{1}\rangle=\sqrt{3}|\phi_{0}\rangle+|\phi_{2}\rangle+|\phi_{3}\rangle$$
$$A_{G_2}|\phi_{2}\rangle=|\phi_{1}\rangle+2|\phi_{3}\rangle$$
\begin{equation}
A_{G_2}|\phi_{3}\rangle=|\phi_{1}\rangle+2|\phi_{2}\rangle+2|\phi_{3}\rangle
\end{equation}
So, the adjacency matrix on strata basis is
\begin{equation}
A_{G_1}=\left(
\begin{array}{cccc}
  0 & \sqrt{3} & 0 & 0 \\
  \sqrt{3} & 2 & 1 & 1 \\
  0 & 1 & 0 & 2 \\
  0 & 1 & 2 & 0 \\
\end{array}
\right)
\end{equation}
\begin{equation}
A_{G_2}=\left(
\begin{array}{cccc}
  0 & \sqrt{3} & 0 & 0 \\
  \sqrt{3} & 0 & 1 & 1 \\
  0 & 1 & 0 & 2 \\
  0 & 1 & 2 & 2 \\
\end{array}\right)
\end{equation}
The adjacency matrices of these two graphs in the first stratification basis are different, so the amplitudes of single particle quantum walk are different for two nonisomorph graphs. Therefore the single particle quantum walk can distinguish graph nonisomorphism.

\textit{Example II:}

Two cospectral nonisomorph graphs $H_1$ and $H_2$ are shown in Fig (3). They have $12$ vertices and $33$ edges. The degree distribution of two graphs are $8,8,8,8,8,8,3,3,3,3,3,3$.

\setlength{\unitlength}{0.75cm}
\begin{picture}(6,8)
\linethickness{0.075mm}

\put(2,7){$(a)H_1$}
\put(12,7){$(b)H_2$}

\put(1.5,1){$6$}
\put(1.5,2){$5$}
\put(1.5,3){$4$}
\put(1.5,4){$3$}
\put(1.5,5){$2$}
\put(1.5,6){$1$}

\put(4.3,1){$12$}
\put(4.3,2){$11$}
\put(4.3,3){$10$}
\put(4.3,4){$9$}
\put(4.3,5){$8$}
\put(4.3,6){$7$}

\put(2,2){\circle*{0.2}}
\put(2,3){\circle*{0.2}}
\put(2,4){\circle*{0.2}}
\put(2,5){\circle*{0.2}}
\put(2,1){\circle*{0.2}}
\put(2,6){\circle*{0.2}}

\put(4,1){\circle*{0.2}}
\put(4,2){\circle*{0.2}}
\put(4,3){\circle*{0.2}}
\put(4,4){\circle*{0.2}}
\put(4,5){\circle*{0.2}}
\put(4,6){\circle*{0.2}}

\put(2,6){\line(1,0){2}}
\put(2,6){\line(2,-1){2}}
\put(2,6){\line(1,-1){2}}
\put(2,5){\line(1,0){2}}
\put(2,5){\line(2,1){2}}
\put(2,5){\line(2,-1){2}}
\put(2,4){\line(1,1){2}}
\put(2,4){\line(2,-1){2}}
\put(2,4){\line(1,-1){2}}
\put(2,3){\line(1,1){2}}
\put(2,3){\line(2,-1){2}}
\put(2,3){\line(1,-1){2}}
\put(2,2){\line(1,1){2}}
\put(2,2){\line(2,1){2}}
\put(2,2){\line(2,-1){2}}
\put(2,1){\line(1,1){2}}
\put(2,1){\line(2,1){2}}
\put(2,1){\line(1,0){2}}
\put(2,1){\line(0,1){5}}
\put(2,5){\oval(2.8,2)[l]}
\put(2,4.5){\oval(3.2,3)[l]}
\put(2,4){\oval(3.6,4)[l]}
\put(2,3.5){\oval(4,5)[l]}
\put(2,4){\oval(1.6,2)[l]}
\put(2,3.5){\oval(2,3)[l]}
\put(2,3){\oval(2.4,4)[l]}
\put(2,3){\oval(0.8,2)[l]}
\put(2,2.5){\oval(1.2,3)[l]}
\put(2,2){\oval(0.4,2)[l]}

\put(11.5,1){$6$}
\put(11.5,2){$5$}
\put(11.5,3){$4$}
\put(11.5,4){$3$}
\put(11.5,5){$2$}
\put(11.5,6){$1$}

\put(14.3,1){$12$}
\put(14.3,2){$11$}
\put(14.3,3){$10$}
\put(14.3,4){$9$}
\put(14.3,5){$8$}
\put(14.3,6){$7$}

\put(12,1){\circle*{0.2}}
\put(12,2){\circle*{0.2}}
\put(12,3){\circle*{0.2}}
\put(12,4){\circle*{0.2}}
\put(12,5){\circle*{0.2}}
\put(12,6){\circle*{0.2}}

\put(14,1){\circle*{0.2}}
\put(14,2){\circle*{0.2}}
\put(14,3){\circle*{0.2}}
\put(14,4){\circle*{0.2}}
\put(14,5){\circle*{0.2}}
\put(14,6){\circle*{0.2}}

\put(12,6){\line(1,0){2}}
\put(12,6){\line(2,-1){2}}
\put(12,6){\line(1,-1){2}}
\put(12,5){\line(1,0){2}}
\put(12,5){\line(2,1){2}}
\put(12,5){\line(1,-1){2}}
\put(12,4){\line(1,1){2}}
\put(12,4){\line(2,1){2}}
\put(12,4){\line(1,-1){2}}
\put(12,3){\line(2,1){2}}
\put(12,3){\line(2,-1){2}}
\put(12,3){\line(1,-1){2}}
\put(12,2){\line(1,1){2}}
\put(12,2){\line(2,1){2}}
\put(12,2){\line(2,-1){2}}
\put(12,1){\line(1,1){2}}
\put(12,1){\line(2,1){2}}
\put(12,1){\line(1,0){2}}
\put(12,1){\line(0,1){5}}
\put(12,5){\oval(2.8,2)[l]}
\put(12,4.5){\oval(3.2,3)[l]}
\put(12,4){\oval(3.6,4)[l]}
\put(12,3.5){\oval(4,5)[l]}
\put(12,4){\oval(1.6,2)[l]}
\put(12,3.5){\oval(2,3)[l]}
\put(12,3){\oval(2.4,4)[l]}
\put(12,3){\oval(0.8,2)[l]}
\put(12,2.5){\oval(1.2,3)[l]}
\put(12,2){\oval(0.4,2)[l]}

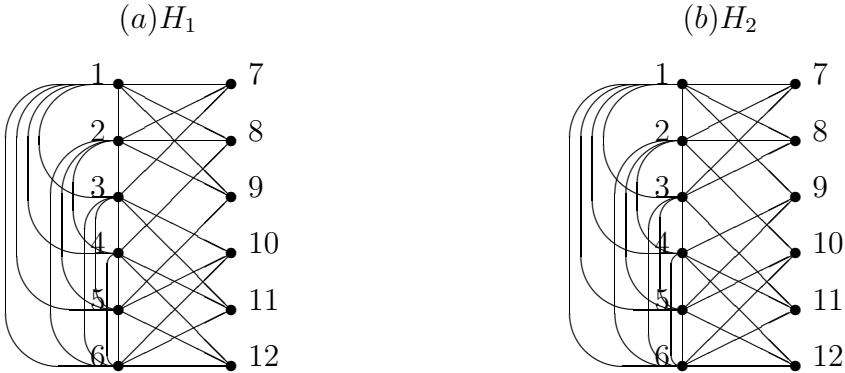
\put(1,0){\footnotesize FIG III: A pair of nonisomorphic cospectral graphs:$(a):H_1$ in the left and the $(b):H_2$ in the right.}
\put(1,-1){\footnotesize Single particle quantum walk can distinguish these two graphs.}
\end{picture}
\newline
\newline
\newline
The stratification basis are defined in the graph $H_1$ as
following
$$|\phi_{0}\rangle=|6\rangle$$
$$|\phi_{1}\rangle=\frac{1}{\sqrt{3}}(|10\rangle+|11\rangle+|12\rangle)$$
$$|\phi_{2}\rangle=\frac{1}{\sqrt{3}}(|3\rangle+|4\rangle+|5\rangle)$$
$$|\phi_{3}\rangle=\frac{1}{\sqrt{3}}(|7\rangle+|8\rangle+|9\rangle)$$
\begin{equation}
|\phi_{4}\rangle=\frac{1}{\sqrt{2}}(|1\rangle+|2\rangle)
\end{equation}
So
$$A_{H_1}|\phi_{0}\rangle=\sqrt{3}|\phi_{1}\rangle+\sqrt{2}|\phi_{4}\rangle+\sqrt{3}|\phi_{2}\rangle$$
$$A_{H_1}|\phi_{1}\rangle=\sqrt{3}|\phi_{0}\rangle+2|\phi_{2}\rangle$$
$$A_{H_1}|\phi_{2}\rangle=\sqrt{3}|\phi_{0}\rangle+2|\phi_{1}\rangle+2|\phi_{2}\rangle+|\phi_{3}\rangle+\sqrt{6}|\phi_{4}\rangle$$
$$A_{H_1}|\phi_{3}\rangle=\sqrt{6}|\phi_{4}\rangle+|\phi_{2}\rangle$$
\begin{equation}
A_{H_1}|\phi_{4}\rangle=\sqrt{2}|\phi_{0}\rangle+\sqrt{6}|\phi_{2}\rangle+\sqrt{6}|\phi_{3}\rangle+|\phi_{4}\rangle
\end{equation}

\begin{equation}
A_{H_1}=\left(\begin{array}{ccccc}
  0 & \sqrt{3} & \sqrt{3}& 0 & \sqrt{2} \\
  \sqrt{3} & 0 & 2 & 0 & 0 \\
  \sqrt{3} & 2 & 2 & 1 & \sqrt{6} \\
  0 & 0 & 1 & 0 & \sqrt{6} \\
  \sqrt{2} & 0 & \sqrt{6} & \sqrt{6} & 1 \\
\end{array}\right)
\end{equation}
The stratification basis are defined in the graph $H_2$ as
following
$$|\phi_{0}\rangle=|12\rangle$$
$$|\phi_{1}\rangle=\frac{1}{\sqrt{3}}(|4\rangle+|5\rangle+|6\rangle)$$
$$|\phi_{2}\rangle=\frac{1}{\sqrt{3}}(|9\rangle+|10\rangle+|11\rangle)$$
$$|\phi_{3}\rangle=\frac{1}{\sqrt{3}}(|1\rangle+|2\rangle+|3\rangle)$$
\begin{equation}
|\phi_{4}\rangle=\frac{1}{\sqrt{2}}(|7\rangle+|8\rangle)
\end{equation}
So
$$A_{H_2}|\phi_{0}\rangle=\sqrt{3}|\phi_{1}\rangle$$
$$A_{H_2}|\phi_{1}\rangle=\sqrt{3}|\phi_{0}\rangle+2|\phi_{1}\rangle+2|\phi_{2}\rangle+3|\phi_{3}\rangle$$
$$A_{H_2}|\phi_{2}\rangle=2|\phi_{1}\rangle+|\phi_{3}\rangle$$
$$A_{H_2}|\phi_{3}\rangle=3|\phi_{1}\rangle+|\phi_{2}\rangle+2|\phi_{3}\rangle+\sqrt{6}|\phi_{4}\rangle$$
\begin{equation}
A_{H_2}|\phi_{4}\rangle=\sqrt{6}|\phi_{3}\rangle
\end{equation}

\begin{equation}
A_{H_2}=\left(\begin{array}{ccccc}
  0 & \sqrt{3} & 0& 0 & 0 \\
  \sqrt{3} & 2 & 2 & 3 & 0 \\
  0 & 2 & 0 & 1 & 0 \\
  0 & 3 & 1 & 2 & \sqrt{6} \\
  0 & 0 & 0 & \sqrt{6} & 0 \\
\end{array}\right)
\end{equation}

The adjacency matrices of these two graphs in the first stratification basis are different, so the amplitudes of single particle quantum walk are different for two nonisomorph graphs. Therefore the single particle quantum walk can distinguish graph nonisomorphism.

\textit{Example III:}

Two graphs $M_1$ and $M_2$ in the Fig (4) are cospectral and nonisomorph. They have $13$ vertices and $15$ edges. The degree distribution of two graphs are $3,3,3,3,2,2,2,3,3,3,1,1,1$.

\setlength{\unitlength}{0.75cm}
\begin{picture}(6,8)
\linethickness{0.075mm}
\put(3.7,7){$(a)M_1$}
\put(13.7,7){$(b)M_2$}

\put(1.5,2){$4$}
\put(1.5,3){$3$}
\put(1.5,4){$2$}
\put(1.5,5){$1$}

\put(4.3,1.1){$10$}
\put(4.3,2.1){$9$}
\put(4.3,3.1){$8$}
\put(4.3,4){$7$}
\put(4.3,5){$6$}
\put(4.3,6){$5$}

\put(6.3,1){$11$}
\put(6.3,2){$12$}
\put(6.3,3){$13$}

\put(2,2){\circle*{0.2}}
\put(2,3){\circle*{0.2}}
\put(2,4){\circle*{0.2}}
\put(2,5){\circle*{0.2}}

\put(4,1){\circle*{0.2}}
\put(4,2){\circle*{0.2}}
\put(4,3){\circle*{0.2}}
\put(4,4){\circle*{0.2}}
\put(4,5){\circle*{0.2}}
\put(4,6){\circle*{0.2}}

\put(6,1){\circle*{0.2}}
\put(6,2){\circle*{0.2}}
\put(6,3){\circle*{0.2}}

\put(2,5){\line(2,1){2}}
\put(2,5){\line(1,0){2}}
\put(2,5){\line(2,-1){2}}
\put(2,4){\line(1,1){2}}
\put(2,4){\line(2,-1){2}}
\put(2,4){\line(1,-1){2}}
\put(2,3){\line(1,1){2}}
\put(2,3){\line(1,0){2}}
\put(2,3){\line(1,-1){2}}
\put(2,2){\line(1,1){2}}
\put(2,2){\line(1,0){2}}
\put(2,2){\line(2,-1){2}}
\put(4,1){\line(1,0){2}}
\put(4,2){\line(1,0){2}}
\put(4,3){\line(1,0){2}}

\put(11.5,2){$4$}
\put(11.5,3){$3$}
\put(11.5,4){$2$}
\put(11.5,5){$1$}

\put(14.3,1.1){$10$}
\put(14.3,2.1){$9$}
\put(14.3,3.1){$8$}
\put(14.3,4){$7$}
\put(14.3,5){$6$}
\put(14.3,6){$5$}

\put(16.3,1){$11$}
\put(16.3,2){$12$}
\put(16.3,3){$13$}

\put(12,2){\circle*{0.2}}
\put(12,3){\circle*{0.2}}
\put(12,4){\circle*{0.2}}
\put(12,5){\circle*{0.2}}

\put(14,1){\circle*{0.2}}
\put(14,2){\circle*{0.2}}
\put(14,3){\circle*{0.2}}
\put(14,4){\circle*{0.2}}
\put(14,5){\circle*{0.2}}
\put(14,6){\circle*{0.2}}

\put(16,1){\circle*{0.2}}
\put(16,2){\circle*{0.2}}
\put(16,3){\circle*{0.2}}

\put(12,5){\line(1,-1){2}}
\put(12,5){\line(2,-3){2}}
\put(12,5){\line(1,-2){2}}
\put(12,4){\line(2,1){2}}
\put(12,4){\line(1,0){2}}
\put(12,4){\line(2,-3){2}}
\put(12,3){\line(2,3){2}}
\put(12,3){\line(2,1){2}}
\put(12,3){\line(2,-1){2}}
\put(12,2){\line(2,1){2}}
\put(12,2){\line(2,3){2}}
\put(12,2){\line(1,2){2}}
\put(14,1){\line(1,0){2}}
\put(14,2){\line(1,0){2}}
\put(14,3){\line(1,0){2}}
\put(1,0){\footnotesize FIG IV: A pair of nonisomorphic cospectral graphs.$(a):M_1$ and $(b):M_2$}
\put(1,-1){\footnotesize Single particle quantum walk can distinguish these two graphs.}
\end{picture}
\newline
\newline
\newline
The stratification basis are defined in the graph $M_1$ as
following
$$|\phi_{0}\rangle=|1\rangle$$
$$|\phi_{1}\rangle=\frac{1}{\sqrt{3}}(|5\rangle+|6\rangle+|7\rangle)$$
$$|\phi_{2}\rangle=\frac{1}{\sqrt{3}}(|2\rangle+|3\rangle+|4\rangle)$$
$$|\phi_{3}\rangle=\frac{1}{\sqrt{3}}(|8\rangle+|9\rangle+|10\rangle)$$
\begin{equation}
|\phi_{4}\rangle=\frac{1}{\sqrt{3}}(|11\rangle+|12\rangle+|13\rangle)
\end{equation}
So
$$A_{M_1}|\phi_{0}\rangle=\sqrt{3}|\phi_{1}\rangle$$
$$A_{M_1}|\phi_{1}\rangle=\sqrt{3}|\phi_{0}\rangle+|\phi_{2}\rangle$$
$$A_{M_1}|\phi_{2}\rangle=|\phi_{1}\rangle+2|\phi_{3}\rangle$$
$$A_{M_1}|\phi_{3}\rangle=|\phi_{4}\rangle+2|\phi_{2}\rangle$$
\begin{equation}
A_{M_1}|\phi_{4}\rangle=|\phi_{3}\rangle
\end{equation}

\begin{equation}
A_{M_1}=\left(\begin{array}{ccccc}
  0 & \sqrt{3} & 0& 0 & 0 \\
  \sqrt{3} & 0 & 1 & 0 & 0 \\
  0 & 1 & 0 & 2 & 0 \\
  0 & 0 & 2 & 0 & 1 \\
  0 & 0 & 0 & 1 & 0 \\
\end{array}\right)
\end{equation}
The stratification basis are defined in the graph $M_2$ as
following
$$|\phi_{0}\rangle=|1\rangle$$
$$|\phi_{1}\rangle=\frac{1}{\sqrt{3}}(|8\rangle+|9\rangle+|10\rangle)$$
$$|\phi_{2}\rangle=\frac{1}{\sqrt{3}}(|2\rangle+|3\rangle+|4\rangle)$$
$$|\phi_{3}\rangle=\frac{1}{\sqrt{3}}(|5\rangle+|6\rangle+|7\rangle)$$
\begin{equation}
|\phi_{4}\rangle=\frac{1}{\sqrt{3}}(|11\rangle+|12\rangle+|13\rangle)
\end{equation}
So
$$A_{M_2}|\phi_{0}\rangle=\sqrt{3}|\phi_{1}\rangle$$
$$A_{M_2}|\phi_{1}\rangle=\sqrt{3}|\phi_{0}\rangle+|\phi_{2}\rangle+|\phi_{4}\rangle$$
$$A_{M_2}|\phi_{2}\rangle=|\phi_{1}\rangle+2|\phi_{3}\rangle$$
$$A_{M_2}|\phi_{3}\rangle=2|\phi_{2}\rangle$$
\begin{equation}
A_{M_2}|\phi_{4}\rangle=|\phi_{1}\rangle
\end{equation}

\begin{equation}
A_{M_2}=\left(\begin{array}{ccccc}
  0 & \sqrt{3} & 0& 0 & 0 \\
  \sqrt{3} & 0 & 1 & 0 & 1 \\
  0 & 1 & 0 & 2 & 0 \\
  0 & 0 & 2 & 0 & 0 \\
  0 & 1 & 0 & 0 & 0 \\
\end{array}\right)
\end{equation}

The adjacency matrices of these two graphs in the first stratification basis are different, so the amplitudes of single particle quantum walk are different for two nonisomorph graphs. Therefore the single particle quantum walk can distinguish graph nonisomorphism.

\section{Investigation of graph isomorphism problem via entanglement entropy in some cospectral graphs}

In this section, we give some examples of Cospectral nonisomorph graphs. Then we rewrite the adjacency matrices of graphs in the stratification basis. In these examples, the new adjacency matrices of graphs are identical. So the single particle quantum walk can not distinguish two nonisomorph graphs. We use the entropy of entanglement to distiguish nonisomorph graphs in this paper.

\textit{example I}:

Two cospectral non-isomorph graphs ($(a):$4-cube graph and $(b):$ Hoffmann graph) are shown in Fig (5). They have $16$ vertices and $32$ edges. Both of them are 4-regular graphs.

\setlength{\unitlength}{0.75cm}
\begin{picture}(6,6)
\linethickness{0.075mm}

\put(4.8,4.8){$(a)$}
\put(16.8,4.8){$(b)$}

\put(4.7,-0.2){$16$}
\put(1.7,0.8){$12$}
\put(3.7,0.8){$13$}
\put(5.7,0.8){$14$}
\put(7.7,0.8){$15$}

\put(-0.2,1.8){$6$}
\put(1.8,1.8){$7$}
\put(3.8,1.8){$8$}
\put(5.8,1.8){$9$}
\put(7.7,1.8){$10$}
\put(9.7,1.8){$11$}

\put(1.8,2.8){$2$}
\put(3.8,2.8){$3$}
\put(5.8,2.8){$4$}
\put(7.8,2.8){$5$}

\put(4.8,3.8){$1$}

\put(5,0){\circle{0.8}}

\put(2,1){\circle{0.8}}
\put(4,1){\circle{0.8}}
\put(6,1){\circle{0.8}}
\put(8,1){\circle{0.8}}

\put(0,2){\circle{0.8}}
\put(2,2){\circle{0.8}}
\put(4,2){\circle{0.8}}
\put(6,2){\circle{0.8}}
\put(8,2){\circle{0.8}}
\put(10,2){\circle{0.8}}

\put(2,3){\circle{0.8}}
\put(4,3){\circle{0.8}}
\put(6,3){\circle{0.8}}
\put(8,3){\circle{0.8}}

\put(5,4){\circle{0.8}}

\put(2,1.4){\line(0,1){0.2}}
\put(2,2.4){\line(0,1){0.2}}
\put(8,1.4){\line(0,1){0.2}}
\put(8,2.4){\line(0,1){0.2}}
\put(4,1.4){\line(0,1){0.2}}
\put(6,2.4){\line(0,1){0.2}}

\put(2.4,3.1){\line(3,1){2.2}}
\put(5.4,3.9){\line(3,-1){2.2}}

\put(4.2,3.3){\line(1,1){0.5}}
\put(5.2,3.7){\line(1,-1){0.5}}

\put(2.4,0.9){\line(3,-1){2.3}}
\put(4.4,0.9){\line(1,-1){0.5}}
\put(5.2,0.3){\line(1,1){0.5}}
\put(5.3,0.2){\line(3,1){2.3}}

\put(0.4,2.1){\line(2,1){1.3}}
\put(0.4,1.9){\line(2,-1){1.3}}
\put(0.4,2.1){\line(4,1){3.2}}
\put(0.4,1.9){\line(4,-1){3.2}}

\put(9.6,2.1){\line(-2,1){1.3}}
\put(9.6,1.9){\line(-2,-1){1.3}}
\put(9.6,2.1){\line(-4,1){3.2}}
\put(9.6,1.9){\line(-4,-1){3.2}}

\put(2.4,2.1){\line(4,1){3.2}}
\put(2.4,1.9){\line(4,-1){3.2}}

\put(4.4,2.1){\line(4,1){3.2}}
\put(4.4,1.9){\line(2,-1){1.3}}
\put(3.6,2.1){\line(-2,1){1.3}}

\put(6.4,1.9){\line(2,-1){1.3}}
\put(5.6,2.1){\line(-2,1){1.3}}
\put(5.6,1.9){\line(-4,-1){3.2}}

\put(7.6,2.1){\line(-4,1){3.2}}
\put(7.6,1.9){\line(-4,-1){3.2}}

\put(17,0){\circle{0.8}}
\put(14,1){\circle{0.8}}
 \put(16,1){\circle{0.8}}
 \put(18,1){\circle{0.8}}
 \put(20,1){\circle{0.8}}
\put(12,2){\circle{0.8}}
\put(14,2){\circle{0.8}}
\put(16,2){\circle{0.8}}
\put(18,2){\circle{0.8}}
\put(20,2){\circle{0.8}}
\put(22,2){\circle{0.8}}
\put(14,3){\circle{0.8}}
 \put(16,3){\circle{0.8}}
 \put(18,3){\circle{0.8}}
 \put(20,3){\circle{0.8}}
\put(17,4){\circle{0.8}}

\put(16.7,-0.2){$16$}
\put(13.7,0.8){$12$}
\put(15.7,0.8){$13$}
\put(17.7,0.8){$14$}
\put(19.7,0.8){$15$}
\put(11.8,1.8){$6$}
\put(13.8,1.8){$7$}
\put(15.8,1.8){$8$}
\put(17.8,1.8){$9$}
\put(19.7,1.8){$10$}
\put(21.7,1.8){$11$}
\put(13.8,2.8){$2$}
\put(15.8,2.8){$3$}
\put(17.8,2.8){$4$}
\put(19.8,2.8){$5$}
\put(16.8,3.8){$1$}

\put(14,1.4){\line(0,1){0.2}}
\put(20,1.4){\line(0,1){0.2}}
\put(14,2.4){\line(0,1){0.2}}
\put(16,2.4){\line(0,1){0.2}}
\put(18,2.4){\line(0,1){0.2}}
\put(20,2.4){\line(0,1){0.2}}

\put(14.4,3.1){\line(3,1){2.2}}
\put(19.6,3.1){\line(-3,1){2.2}}
\put(16.2,3.3){\line(1,1){0.5}}
\put(17.8,3.3){\line(-1,1){0.5}}

\put(16.6,0.1){\line(-3,1){2.3}}
\put(17.4,0.1){\line(3,1){2.3}}
\put(17.2,0.3){\line(1,1){0.5}}
\put(16.8,0.3){\line(-1,1){0.5}}

\put(12.4,2.1){\line(2,1){1.3}}
\put(12.4,1.9){\line(2,-1){1.3}}
\put(12.4,2.1){\line(4,1){3.2}}
\put(12.4,1.9){\line(4,-1){3.2}}

\put(21.6,2.1){\line(-2,1){1.3}}
\put(21.6,1.9){\line(-2,-1){1.3}}
\put(21.6,2.1){\line(-4,1){3.2}}
\put(21.6,1.9){\line(-4,-1){3.2}}

\put(14.4,2.1){\line(4,1){3.2}}
\put(14.4,1.9){\line(4,-1){3.2}}

\put(16.4,1.9){\line(4,-1){3.2}}
\put(16.3,1.8){\line(2,-1){1.3}}
\put(15.6,2.1){\line(-2,1){1.3}}

\put(18.4,2.1){\line(2,1){1.3}}
\put(17.7,1.8){\line(-2,-1){1.3}}
\put(17.6,1.9){\line(-4,-1){3.2}}

\put(19.6,2.1){\line(-4,1){3.2}}
\put(19.6,1.9){\line(-4,-1){3.2}}
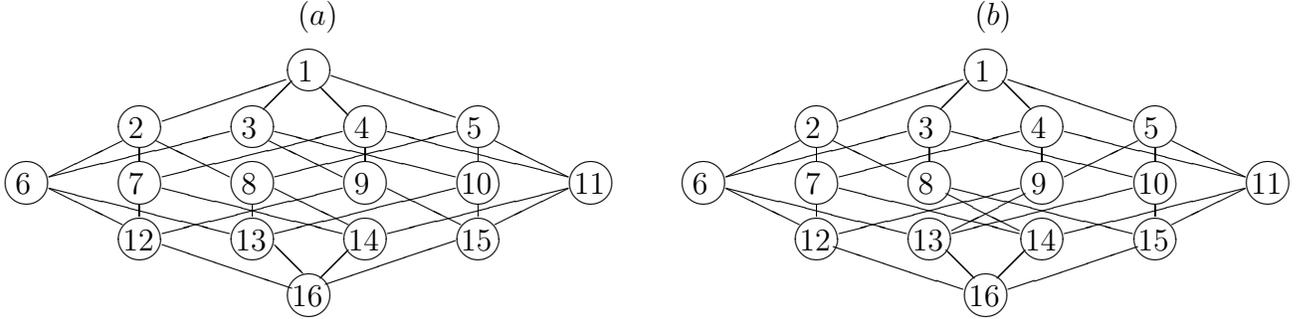
\put(1,-1){\footnotesize FIG V: A pair of nonisomorphic cospectral graphs.$(a):$4-cube and $(b):$Hoffmann graph.}
\put(1,-2){\footnotesize entanglement entropy between strata can distinguish these two graphs.}
\end{picture}
\newline
\newline
\newline
We define the stratification basis as following
$$|\varphi_0\rangle=|1\rangle$$
$$|\varphi_1\rangle=\frac{1}{2}(|2\rangle+|3\rangle+|4\rangle+|5\rangle)$$
$$|\varphi_2\rangle=\frac{1}{\sqrt{6}}(|6\rangle+|7\rangle+|8\rangle+|9\rangle+|10\rangle+|11\rangle)$$
$$|\varphi_3\rangle=\frac{1}{2}(|12\rangle+|13\rangle+|14\rangle+|15\rangle)$$
\begin{equation}
|\varphi_4\rangle=|16\rangle
\end{equation}

So the effect of adjacency matrix on these vectors will be
$$A|\varphi_0\rangle=2|\varphi_1\rangle$$
$$A|\varphi_1\rangle=2|\varphi_0\rangle+\sqrt{6}|\varphi_2\rangle$$
$$A|\varphi_2\rangle=\sqrt{6}|\varphi_1\rangle+\sqrt{6}|\varphi_3\rangle$$
$$A|\varphi_3\rangle=\sqrt{6}|\varphi_2\rangle+2|\varphi_4\rangle$$
\begin{equation}
A|\varphi_4\rangle=2|\varphi_3\rangle
\end{equation}

Then the adjacency matrices of two graphs in the stratification basis, are the same
\begin{equation}
A_1=\left(\begin{array}{ccccc}
          0 & 2 & 0 & 0 & 0\\
          2 & 0 & \sqrt{6} & 0 & 0\\
          0 & \sqrt{6} & 0 & \sqrt{6} & 0\\
          0 & 0 & \sqrt{6} & 0 & 2\\
          0 & 0 & 0 & 2 & 0\\
          \end{array}\right)
\end{equation}

Therefore the amplitudes of single particle quantum walk are identical for these two graphs. So single particle quantum walk fails to distinguish these pairs. Also the entanglement entropy between the vertex of first stratum ($|\varphi_0\rangle=|1\rangle$) and the other vertices can not distinguish these two non-isomorph graphs. But the entanglement entropies between two other parts of above graphs can distinguish these non-isomorph graphs. For examples we separate the vertices of each of these graphs into two subsets. The first subset is the vertices of first and second strata and the other subset is the vertices of third and fourth and fifth strata. Then the entanglement entropy between two subsets are different for two graphs. Therefore the entanglement entropy can distinguish non-isomorph pairs.

\textit{Example II}:

Two non-isomorph graphs in Fig (6) have ten vertices and twenty edges. They are 4-regualr and cospectral.

\setlength{\unitlength}{0.75cm}
\begin{picture}(6,6)
\linethickness{0.075mm}

\put(1.8,2.8){$1$}
\put(5.8,2.8){$2$}
\put(9.8,2.8){$3$}
\put(9.8,-2.2){$4$}
\put(1.8,-2.2){$5$}
\put(7.8,0.8){$6$}
\put(7.8,-0.2){$7$}
\put(5.8,-2.2){$8$}
\put(3.8,-0.2){$9$}
\put(3.7,0.8){$10$}

\put(2,-2){\circle{0.8}}
\put(4,0){\circle{0.8}}
\put(6,-2){\circle{0.8}}
\put(8,0){\circle{0.8}}
\put(10,-2){\circle{0.8}}
\put(4,1){\circle{0.8}}
\put(8,1){\circle{0.8}}
\put(2,3){\circle{0.8}}
\put(6,3){\circle{0.8}}
\put(10,3){\circle{0.8}}

\put(2.4,-2){\line(1,0){3.2}}
\put(6.4,-2){\line(1,0){3.2}}
\put(2,-1.6){\line(0,1){4.2}}
\put(10,-1.6){\line(0,1){4.2}}
\put(2.4,3){\line(1,0){3.2}}
\put(6.4,3){\line(1,0){3.2}}
\put(4,0.4){\line(0,1){0.2}}
\put(8,0.4){\line(0,1){0.2}}

\put(6.4,-1.9){\line(1,1){1.5}}
\put(5.6,-1.9){\line(-1,1){1.5}}
\put(5.6,2.9){\line(-1,-1){1.5}}
\put(6.4,2.9){\line(1,-1){1.5}}

\put(2.4,-1.9){\line(1,1){1.5}}
\put(2.2,-1.7){\line(2,3){1.6}}
\put(9.6,-1.9){\line(-1,1){1.5}}
\put(9.8,-1.7){\line(-2,3){1.6}}

\put(2.4,2.9){\line(1,-1){1.5}}
\put(9.6,2.9){\line(-1,-1){1.5}}
\put(2.4,3){\line(2,-1){5.4}}
\put(9.6,3){\line(-2,-1){5.4}}

\put(11.8,2.8){$1$}
\put(15.8,2.8){$2$}
\put(19.8,2.8){$3$}
\put(19.8,-2.2){$4$}
\put(11.8,-2.2){$5$}
\put(17.8,0.8){$6$}
\put(17.8,-0.2){$7$}
\put(15.8,-2.2){$8$}
\put(13.8,-0.2){$9$}
\put(13.7,0.8){$10$}

\put(12,-2){\circle{0.8}}
\put(14,0){\circle{0.8}}
\put(16,-2){\circle{0.8}}
\put(18,0){\circle{0.8}}
\put(20,-2){\circle{0.8}}
\put(14,1){\circle{0.8}}
\put(18,1){\circle{0.8}}
\put(12,3){\circle{0.8}}
\put(16,3){\circle{0.8}}
\put(20,3){\circle{0.8}}

\put(12.4,-2){\line(1,0){3.2}}
\put(16.4,-2){\line(1,0){3.2}}
\put(12,-1.6){\line(0,1){4.2}}
\put(20,-1.6){\line(0,1){4.2}}
\put(12.4,3){\line(1,0){3.2}}
\put(16.4,3){\line(1,0){3.2}}
\put(14,0.4){\line(0,1){0.2}}
\put(18,0.4){\line(0,1){0.2}}

\put(16.4,-1.9){\line(1,1){1.5}}
\put(15.6,-1.9){\line(-1,1){1.5}}
\put(15.6,2.9){\line(-1,-1){1.5}}
\put(16.4,2.9){\line(1,-1){1.5}}

\put(12.4,-1.9){\line(3,1){5.3}}
\put(12.2,-1.7){\line(2,3){1.6}}
\put(19.6,-1.9){\line(-3,1){5.3}}
\put(19.8,-1.7){\line(-2,3){1.6}}

\put(12.4,2.9){\line(1,-1){1.5}}
\put(19.6,2.9){\line(-1,-1){1.5}}
\put(12,2.6){\line(2,-3){1.65}}
\put(20,2.6){\line(-2,-3){1.65}}
\put(1,-3){\footnotesize FIG VI: A pair of nonisomorphic cospectral graphs.}
\put(1,-4){\footnotesize Entanglement entropy between strata can distinguish these two graphs.}
\end{picture}
\newline
\newline
\newline
We define the stratification basis as following
$$|\varphi_0\rangle=|2\rangle$$
$$|\varphi_1\rangle=\frac{1}{2}(|1\rangle+|3\rangle+|6\rangle+|10\rangle)$$
$$|\varphi_2\rangle=\frac{1}{2}(|4\rangle+|5\rangle+|7\rangle+|9\rangle)$$
\begin{equation}
|\varphi_3\rangle=|8\rangle
\end{equation}

So the effect of adjacency matrix on these vectors will be
$$A|\varphi_0\rangle=2|\varphi_1\rangle$$
$$A|\varphi_1\rangle=2|\varphi_0\rangle+|\varphi_1\rangle+2|\varphi_2\rangle$$
$$A|\varphi_2\rangle=2|\varphi_1\rangle+|\varphi_2\rangle+2|\varphi_3\rangle$$
\begin{equation}
A|\varphi_3\rangle=2|\varphi_2\rangle
\end{equation}

Then the adjacency matrices of two graphs are the same:
\begin{equation}
A_1=\left(\begin{array}{cccc}
          0 & 2 & 0 & 0\\
          2 & 1 & 2 & 0\\
          0 & 2 & 1 & 2\\
          0 & 0 & 2 & 0\\
          \end{array}\right)
\end{equation}

Therefore the amplitudes of single particle quantum walk are identical for these two graphs. So single particle quantum walk fails to distinguish these pairs. Also similar to previous example, the entanglement entropy between the vertex of first stratum ($|\varphi_0\rangle=|2\rangle$) and the other vertices can not distinguish these two non-isomorph graphs. But the entanglement entropies between two other parts of above graphs can distinguish these non-isomorph graphs. For examples we separate the vertices of each of these graphs into two subsets. The first subset is the vertices of first and second strata and the other subset is the vertices of third and fourth strata. Then the entanglement entropy between two subsets are different for two graphs. Therefore the entanglement entropy can distinguish non-isomorph pairs.

\textit{Example III}:

The following two nonisomorph graphs have twelve vertices and twenty four edges. They are 4-regualr and cospectral.

\setlength{\unitlength}{0.75cm}
\begin{picture}(6,12)
\linethickness{0.075mm}

\put(1.5,10){$(a)\Gamma_1$}
\put(11.5,10){$(b)\Gamma_2$}

\put(0.7,6){$1$}
\put(1.9,5.2){$3$}
\put(1.9,6.5){$2$}
\put(2.9,6.2){$4$}
\put(3.9,5.2){$6$}
\put(3.9,6.5){$5$}
\put(4.9,6.2){$7$}
\put(5.9,5.2){$9$}
\put(5.9,6.5){$8$}
\put(6.9,6.2){$10$}
\put(3.2,10.7){$11$}
\put(3.2,1){$12$}

\put(1,6){\circle*{0.2}}
\put(2,5){\circle*{0.2}}
\put(2,7){\circle*{0.2}}
\put(3,6){\circle*{0.2}}
\put(4,5){\circle*{0.2}}
\put(4,7){\circle*{0.2}}
\put(5,6){\circle*{0.2}}
\put(6,5){\circle*{0.2}}
\put(6,7){\circle*{0.2}}
\put(7,6){\circle*{0.2}}
\put(4,1){\circle*{0.2}}
\put(4,11){\circle*{0.2}}

\put(1,6){\line(3,-5){3}}
\put(1,6){\line(3,5){3}}
\put(1,6){\line(1,1){1}}
\put(1,6){\line(1,-1){1}}
\put(2,5){\line(1,1){1}}
\put(2,5){\line(1,0){2}}
\put(2,5){\line(1,-2){2}}
\put(2,7){\line(1,2){2}}
\put(2,7){\line(1,0){2}}
\put(2,7){\line(1,-1){1}}
\put(3,6){\line(1,1){1}}
\put(3,6){\line(1,-1){1}}
\put(4,5){\line(1,0){2}}
\put(4,5){\line(1,1){1}}
\put(4,7){\line(1,-1){1}}
\put(4,7){\line(1,0){2}}
\put(5,6){\line(1,1){1}}
\put(5,6){\line(1,-1){1}}
\put(6,5){\line(1,1){1}}
\put(6,7){\line(1,-1){1}}
\put(4,1){\line(1,2){2}}
\put(4,1){\line(3,5){3}}
\put(4,11){\line(1,-2){2}}
\put(4,11){\line(3,-5){3}}

\put(10.7,6){$1$}
\put(12.1,5.5){$3$}
\put(12.1,6.2){$2$}
\put(12.9,6.2){$4$}
\put(14.1,5.5){$6$}
\put(14.1,6.2){$5$}
\put(14.9,6.2){$7$}
\put(16.1,5.5){$9$}
\put(16.1,6.2){$8$}
\put(16.9,6.2){$10$}
\put(13.2,10.7){$11$}
\put(13.2,1){$12$}

\put(11,6){\circle*{0.2}}
\put(12,5){\circle*{0.2}}
\put(12,7){\circle*{0.2}}
\put(13,6){\circle*{0.2}}
\put(14,5){\circle*{0.2}}
\put(14,7){\circle*{0.2}}
\put(15,6){\circle*{0.2}}
\put(16,5){\circle*{0.2}}
\put(16,7){\circle*{0.2}}
\put(17,6){\circle*{0.2}}
\put(14,1){\circle*{0.2}}
\put(14,11){\circle*{0.2}}

\put(14,7){\oval(4,1)[t]}
\put(14,5){\oval(4,1)[b]}
\put(14,6){\oval(8,10)[r]}

\put(11,6){\line(3,-5){3}}
\put(11,6){\line(3,5){3}}
\put(11,6){\line(1,1){1}}
\put(11,6){\line(1,-1){1}}
\put(12,5){\line(0,1){2}}
\put(12,5){\line(1,1){1}}
\put(12,7){\line(1,-1){1}}
\put(13,6){\line(1,1){1}}
\put(13,6){\line(1,-1){1}}
\put(14,5){\line(1,1){1}}
\put(14,5){\line(0,1){2}}
\put(14,7){\line(1,-1){1}}
\put(14,7){\line(0,1){4}}
\put(15,6){\line(1,1){1}}
\put(15,6){\line(1,-1){1}}
\put(16,5){\line(1,1){1}}
\put(16,5){\line(0,1){2}}
\put(16,7){\line(1,-1){1}}
\put(14,1){\line(0,1){4}}
\put(14,1){\line(3,5){3}}
\put(14,11){\line(3,-5){3}}

\put(1,0){\footnotesize FIG VII: A pair of nonisomorphic cospectral graphs:$(a):\Gamma_1$ and $(b):\Gamma_2$.}
\put(1,-1){\footnotesize The entanglement entropy can distinguish these two graphs.}
\end{picture}
\newline
\newline
\newline
We define the stratification basis as following
$$|\varphi_0\rangle=|1\rangle$$
$$|\varphi_1\rangle=\frac{1}{2}(|2\rangle+|3\rangle+|11\rangle+|12\rangle)$$
$$|\varphi_2\rangle=\frac{1}{2}(|5\rangle+|6\rangle+|8\rangle+|9\rangle)$$
$$|\varphi_3\rangle=\frac{1}{\sqrt{2}}(|4\rangle+|10\rangle)$$
\begin{equation}
|\varphi_4\rangle=|7\rangle
\end{equation}

So the effect of adjacency matrix on these vectors will be
$$A|\varphi_0\rangle=2|\varphi_1\rangle$$
$$A|\varphi_1\rangle=2|\varphi_0\rangle+|\varphi_1\rangle+|\varphi_2\rangle+\sqrt{2}|\varphi_3\rangle$$
$$A|\varphi_2\rangle=|\varphi_1\rangle+|\varphi_2\rangle+\sqrt{2}|\varphi_3\rangle+2|\varphi_4\rangle$$
$$A|\varphi_3\rangle=\sqrt{2}|\varphi_1\rangle+\sqrt{2}|\varphi_2\rangle$$
\begin{equation}
A|\varphi_4\rangle=2|\varphi_2\rangle
\end{equation}

Then the adjacency matrices of two graphs in these stratification basis are the same:
\begin{equation}
A_1=\left(\begin{array}{ccccc}
          0 & 2 & 0 & 0        & 0\\
          2 & 1 & 1 & \sqrt{2} & 0\\
          0 & 1 & 1 & \sqrt{2} & 2\\
          0 & \sqrt{2} & \sqrt{2} & 0 & 0\\
          0 & 0 & 2 & 0 & 0\\
          \end{array}\right)
\end{equation}

Therefore the amplitudes of single particle quantum walk are identical for these two graphs. So single particle quantum walk fails to distinguish these pairs. Also similar to previous examples of this section, the entanglement entropy between the vertex of first stratum ($|\varphi_0\rangle=|1\rangle$) and the other vertices can not distinguish these two non-isomorph graphs. But the entanglement entropies between two other parts of above graphs can distinguish these non-isomorph graphs. For examples we separate the vertices of each of these graphs into two subsets. The first subset is the vertices of first and second strata and the other subset is the vertices of third, fourth and fifth strata. After calculations based on Appendix A, we concluded that the entanglement entropy between two subsets are different for two graphs. Therefore the entanglement entropy can distinguish non-isomorph pairs.

\section{Conclusion}
We investigated the graph
isomorphism problem, in which one wishes to determine whether two graphs are isomorphic. In two non-isomorph cospectral graphs $G_4(n,n+2)$ and $G_5(n,n+2)$, we used $n$-particle quantum walk to distinguish these two graphs. It was performed by using the antisymmetric fermionic basis. The adjacency matrices of graphs was written in these new basis. It was different for pairs of non-isomorph graphs, so the $n$-particle quantum walk could detect non-isomorph pairs. Also in two other similar cases $T_4(n,n+2)$ and $T_5(n,n+2)$ the $n$-particle quantum walk could detect these graphs. Then the entanglement entropy was used for GI problem in these graphs. It was shown that both of $n$-particle quantum walk and entanglement entropy can detect non-isomorph pairs of $G_4(n,n+2)$ and $G_5(n,n+2)$.

In some examples of non-isomorph cospectral graphs, we show that the single particle quantum walk can detect non-isomorphism. Finally we give some other pairs of non-isomorph examples which their adjacency matrices in the stratification basis are the same. So the single particle quantum walk fails to distinguish pairs of non-isomorph graphs but the entanglement entropy between strata are different in these graphs. Therefore the entanglement entropy can be used for GI problem in these graphs.

One expect that the quantum walk in antisymmetric basis be able to distinguish some other kinds of graphs. Also it seems that the entanglement entropy is a powerfull tool for detecting non-isomorph graphs.
\section*{Appendix}
\appendix

\section{Entanglement entropy between two parts of a network}
we want to use a method to quantify the entanglement entropy between two arbitrary parts of a network. This process is completely explained in [23].
First we divide the potential matrix of the system into two parts, So
the potential matrix ($I+2gL$) can be written in the form
\begin{equation}\label{pot2}
\left(\begin{array}{cc}
          A& B\\
            B^T& C\\
          \end{array}\right)
\end{equation}
where the size of block $A$ is $m\times m$, $C$ is
$(N-m)\times(N-m)$ and $B$ is $m\times(N-m)$. We assume that the
vector $X$ be decomposed of two sets $X,Y$ (i.e.,
$X=(X|Y)=(x_1,x_2, ..., x_m,y_1,y_2,...,y_{N-m})$).

We know that any local operation dosen't change the entanglement
between the nodes, so we apply some of these operations to
calculate the entanglement entropy for different lattices easily.
First we want to diagonalize the blocks $A$ and $C$, to this aim
we apply the local rotations $O_A$ and $O_C$ to the matrix of
(\ref{pot2}), resulting as:

\begin{equation}
\psi(\widehat{x},\widehat{y})=A_g
exp(-\frac{1}{2}(\widehat{x}\quad \quad
\widehat{y})\left(\begin{array}{cc}
          D_A& \hat{B}\\
            \hat{B}^T & D_C\\
          \end{array}\right)\left(\begin{array}{c}
          \widehat{x}\\
            \widehat{y}\\
          \end{array}\right))
\end{equation}
where $\widehat{x}=O^{\dagger}_A x$, $\widehat{y}=O_C^{\dagger} y$, $\hat{B}=O^{\dagger}_A B O_C$, $D_A=O_A^{\dagger}AO_A$ and $D_C=O_C^{\dagger}CO_C$.

In the next stage, the blocks $D_A$ and $D_C$ can be transformed
to Identity matrices by rescaling the variables $\widehat{x}$ and
$\widehat{y}$ as  $\widetilde{x}=D_x^{1/2}\widehat{x}$ and $\widetilde{y}=D_y^{1/2}\widehat{y}$.
So the ground state wave function is transformed to
\begin{equation}
\psi(\widetilde{x},\widetilde{y})=A_gexp(-\frac{1}{2}(\widetilde{x}\quad
\quad \widetilde{y})\left(\begin{array}{cc}
          I& \widetilde{B}\\
            \widetilde{B}^T& I\\
          \end{array}\right)\left(\begin{array}{c}
          \widetilde{x}\\
            \widetilde{y}\\
          \end{array}\right))
\end{equation}
where $\widetilde{B}=D_x^{-1/2}\hat{B}D_y^{-1/2}$

In the third stage, we should calculate the singular value
decomposition of matrix $\widetilde{B}$ as $U\widetilde{B}V^{\dagger}=D_B$. So the variables become $U\widetilde{x}=x'$ and $V\widetilde{y}=y'$
The ground state wave function after this local operation is:

\begin{equation}
\psi(x',y')=A_gexp(-\frac{1}{2}(x'\quad \quad
y')\left(\begin{array}{cc}
          I& U\widetilde{B}V^{\dagger}\\
            V\widetilde{B}^TU^{\dagger}& I\\
          \end{array}\right)\left(\begin{array}{c}
          x'\\
            y'\\
          \end{array}\right))
\end{equation}
The singular value decomposition(SVD), transforms matrix
$\widetilde{B}$ to the diagonal matrix $D_B$ as
$$U\widetilde{B}V^{\dagger}=\left(\begin{array}{cccc}
          d_1& 0&\ldots & 0\\
          0&d_2&\ldots &0\\
          \vdots &\vdots &\ddots &\vdots\\
          0&0&\ldots&d_m\\
          \end{array}\right)$$

Then the final form of wave function is:
$$\psi(q^{x}_1,q^{x}_2,...,q_m^{x},q_1^y,q_2^y,...,q_{N-m}^y)=$$
$$A_ge^{-\frac{(q_1^x)^2}{2}-\frac{(q_1^y)^2}{2}-d_1q_1^xq_1^y}\times
e^{-\frac{(q_2^x)^2}{2}-\frac{(q_2^y)^2}{2}-d_2q_2^xq_2^y}\times
\ldots \times
e^{-\frac{(q_m^x)^2}{2}-\frac{(q_m^y)^2}{2}-d_mq_m^xq_m^y}$$
\begin{equation}\label{Waveff}
\times e^{-\frac{(q_{m+1}^y)^2}{2}}\times
e^{-\frac{(q_{m+2}^y)^2}{2}}\times \ldots \times
e^{-\frac{(q_{N-m}^y)^2}{2}}
\end{equation}
From above equation, it's clear that the node $q_i^x$ is just
entangled with $q_i^y$, so we can use following identity to
calculate the schmidt number of this wave function,
\begin{equation}
\frac{1}{\pi^{1/2}}exp(-\frac{1+t^2}{2(1-t^2)}((q_i^x)^2+(q_i^y)^2))+\frac{2t}{1-t^2}q_i^x
q_i^y)=(1-t^2)^{1/2}\sum_n t^n \psi_n(q_i^x)\psi_n(q_i^y)
\end{equation}
In order to calculating the entropy, we apply a change of
variable as
$$1-t^2=\frac{2}{\nu+1}$$
$$t^2=\frac{\nu-1}{\nu+1}$$
So the above identity becomes
\begin{equation}\label{identity2}
\frac{1}{\pi^{1/2}}exp(-\frac{\nu}{2}((q_i^x)^2+(q_i^y)^2))+(\nu^2-1)^{1/2}q_i^x
q_i^y)=(\frac{2}{\nu+1})^{1/2}\sum_n (\frac{\nu-1}{\nu+1})^{n/2}
\psi_n(q_i^x)\psi_n(q_i^y)
\end{equation}
and the reduced density matrix is
\begin{equation}
\rho=\frac{2}{\nu+1}\sum_{n}(\frac{\nu-1}{\nu+1})^n |n\rangle
\langle n|
\end{equation}
By considering $p_n=\frac{2}{\nu+1}(\frac{\nu-1}{\nu+1})^n$, the entropy is
\begin{equation}\label{Srho}
S(\rho)=-\sum_n p_n log(p_n)=\frac{\nu +1}{2} log(\frac{\nu +1}{2})-\frac{\nu
-1}{2}log(\frac{\nu -1}{2})
\end{equation}

By comparing the wave function (\ref{Waveff}) and the identity (\ref{identity2}) and
define the scale $\mu^2$, we conclude that $\nu_i=1 \times \mu^2$ and $(\nu_i^2-1)^{1/2}=-d_i \times \mu^2$
After some straightforward calculation we obtain
\begin{equation}\label{nud}
\nu_i=(\frac{1}{1-d_i^2})^{1/2}
\end{equation}
By above discussion we conclude that
$$e^{-\frac{(q_i^x)^2}{2}-\frac{(q_i^y)^2}{2}-d_iq_i^xq_i^y}=\sum_n \lambda_{i,n}\psi_n(q^x_i)\psi_n(q^y_i)$$
where
$\lambda_{i,n}=(\frac{2}{\nu_i+1})^{1/2}(\frac{\nu_i-1}{\nu_i+1})^{n/2}$.

Therefore the entropy of each part can be written
$$S(\rho_i)=\frac{\nu_i +1}{2} log(\frac{\nu_i +1}{2})-\frac{\nu_i
-1}{2}log(\frac{\nu_i -1}{2})$$
So the total entropy is
\begin{equation}
S(\rho)=\sum_i S(\rho_i)
\end{equation}

\section{Generalized Schur complement method}
We want to calculate the entanglement entropy between two parts in an arbitrary graph. Suppose there are $m$ ($n$) nodes in the first (second) part. There are $m_1$ ($n_1$) nodes in the first (second) part which are not connected to the nodes of other part. one can separate the nodes of each part $m,n$ into two subsets, so there are four subsets which have
$m_1,m_2,n_2,n_1$ nodes, respectively.
The potential matrix of the system is:
\begin{equation}
V=I+2gL=\left(\begin{array}{cccc}
          V_{11}& V_{12} & 0 &0\\
            V_{12}^T & V_{22} & V_{23} & 0 \\
            0 & V_{23}^T & V_{33} & V_{34}\\
            0 & 0 & V_{34}^T & V_{44} \\
          \end{array}\right)
\end{equation}

Then by using the Generalized Schur complement theorem, we can
write

$$\left(\begin{array}{cccc}
          V_{11}& V_{12} & 0 &0\\
            V_{12}^T & V_{22} & V_{23} & 0 \\
            0 & V_{23}^T & V_{33} & V_{34}\\
            0 & 0 & V_{34}^T & V_{44} \\
          \end{array}\right)=$$
\begin{equation}
          =\left(\begin{array}{cccc}
          I_{m_1}& 0 & 0 & 0\\
            V_{12}^TV_{11}^{-1}& I_{m_2} & 0 & 0 \\
           0 & 0 & I_{n_2}& V_{34}V_{44}^{-1}\\
           0 & 0 & 0 & I_{n_1}\\
          \end{array}\right)\left(\begin{array}{cccc}
          V_{11}& 0 & 0 & 0 \\
           0 & \widetilde{V}_{22} & V_{23} & 0\\
            0 & V_{23}^T & \widetilde{V}_{33}  & 0\\
            0 & 0 & 0 & V_{44} \\
          \end{array}\right)\left(\begin{array}{cccc}
          I_{m_1}& V_{11}^{-1}V_{12} & 0 & 0 \\
            0& I_{m_2}& 0 & 0 \\
            0 & 0 & I_{n_2} & 0 \\
            0 & 0 & V_{44}^{-1}V_{34}^T & I_{n_1} \\
          \end{array}\right)
\end{equation}
Where $\widetilde{V}_{22}=V_{22}-V_{12}^T V_{11}^{-1}V_{12}$ and
$\widetilde{V}_{33}=V_{33}-V_{34}V_{44}^{-1}V_{34}^T$.

\end{document}